\documentclass[authoryear]{elsarticle}
\usepackage{graphicx,subfig}
\usepackage[top=1.2in, bottom=1.35in, left=1.4in, right=1.4in]{geometry}
\usepackage{amsmath,amsfonts,amssymb,amsbsy,amscd,amsgen,amsthm}
\usepackage{dsfont}
\usepackage[]{natbib}
\usepackage{color}
\usepackage{appendix}
\usepackage[colorlinks]{hyperref}
\AtBeginDocument{\hypersetup{citecolor=black}}
\usepackage{comment}
\usepackage{enumerate}
\usepackage{dcolumn}
\usepackage{gensymb}
\usepackage[ruled,vlined,norelsize]{algorithm2e}

\newcommand{\id}{\mathrm d}
\newcommand{\vc}{\pmb}

\newcommand{\Amax}{A_{max}}
\newcommand{\Aext}{A_{ext}}
\newcommand{\Pext}{P_{ext}}

\begin{document}

\begin{frontmatter}
\title{Reduced-order prediction of rogue waves in two-dimensional deep-water waves}
\author{Mohammad Farazmand}
\author{Themistoklis P.  Sapsis\corref{cor2}}
\cortext[cor2]{Corresponding author:
\href{mailto:sapsis@mit.edu}{sapsis@mit.edu},
Tel: (617) 324-7508, Fax: (617) 253-8689}
\address{Department of Mechanical Engineering,
Massachusetts Institute of Technology, \\
77 Massachusetts Ave., Cambridge, MA 02139}

\begin{abstract}
We consider the problem of large wave prediction in two-dimensional water waves. Such waves form due to the synergistic effect of dispersive mixing of smaller wave groups and the action of localized nonlinear wave interactions that leads to focusing. Instead of a direct simulation approach, we rely on the  decomposition of the  wave field into a discrete set of localized wave groups with optimal length scales and amplitudes. Due to the short-term character of the prediction, these wave groups do not interact and therefore their dynamics can be characterized individually. Using direct numerical simulations of the governing envelope equations we precompute the expected maximum elevation for each of those wave groups. The combination of the wave field decomposition algorithm, which provides information about the statistics of the system, and the precomputed map for the expected wave group elevation, which  encodes dynamical information, allows (i) for  understanding of how the probability of occurrence of rogue waves changes as the spectrum parameters vary, (ii) the computation of a critical length scale characterizing wave groups with high probability of evolving to rogue waves, and (iii) the formulation of a robust and parsimonious reduced-order prediction scheme for large waves. We assess the validity of this scheme in several cases of ocean wave spectra. 
\end{abstract}


\end{frontmatter}

\section{Introduction}
Rogue waves refer to extremely large oceanic surface waves.
As a result of their devastating impact on marine systems, such as ships and offshore platforms, 
rogue waves have been the subject of numerous theoretical, experimental and numerical 
studies~\citep{dysthe08,chabchoub2011,Onorato13}. 
Most studies concern the frequency and statistics of rogue wave occurrence for a given sea 
state (see, e.g.,~\cite{longuet1952,tayfun1980,forristall2000,janssen03,xiao2013}). 
It is, however, often desirable to know, for a given ocean area, 
\emph{if}, \emph{when} and \emph{where} a rogue wave may occur in the future. 

These questions can in principle be addressed by numerically solving the 
appropriate hydrodynamic equations~\citep{mei2005, dommermuth1987, Clauss2014}. Apart from its high computational cost,
this direct approach requires a well-resolved state of the fluid velocity field 
and its free surface elevation as initial conditions. Thanks to recent developments, 
real-time and reliable measurement of
sea surface elevation is feasible (see e.g.~\cite{nieto04,story11,fu_2011,borge13,Trillo16}). 
But the well-resolved measurement of fluid velocity field remains out of reach.

An alternative approach for short-term prediction of the
wave field is based on numerically solving 
the so-called envelope equations, which approximate the evolution of the wave envelope to a 
reasonable accuracy~\citep{zakharov68,dysthe79,trulsen96}. While less 
expensive than the full hydrodynamic equations, solving the envelope equations is still computationally 
formidable for real-time forecast of extreme waves. 

As a result, several attempts have been made to 
devise reliable, reduced-order methods for short-time forecast of water wave evolution.
\cite{Adcock12}, for instance, approximate nonlinear evolution of localized wave groups 
with an exact breather-like solution of the linear Schr\" odinger equation. To account
for the nonlinear effects, they allow the parameters of
the breather-like solution to vary in time such that particular invariants (energy and the
Hamiltonian) of the nonlinear Schr\" odinger equation (NLS) are preserved over time. 
\cite{ruban2015b} takes a similar approach by substituting a Gaussian ansatz
into the Lagrangian functional associated with the NLS equation. The time evolution of the parameters
are determined such that the solution satisfies a least-action principle (also see~\cite{ruban2015}). 

The resulting wave groups from \cite{Adcock12} and \cite{ruban2015} do not necessarily satisfy the underlying envelope equation (i.e., the NLS equation). 
Furthermore, the reduced-order method of~\cite{ruban2015} relies heavily on the Lagrangian formulation of the NLS equation.
As such, it is not immediately applicable to the more realistic, higher-order envelope equations, such as the 
modified NLS (MNLS) equation of~\cite{dysthe79}, whose Lagrangian formulation is unavailable
(see~\cite{gramstad2011} and \cite{Craig2012} for the Hamiltonian formulation of the MNLS equation).

To avoid these drawbacks,~\cite{cousins16, cousins14} take an intermediate approach. They also consider the evolution
of parametric wave groups but allow the wave group to evolve under the full non-linear evolution equation by imposing energy conservation~\citep{cousins15}. The analysis resulted in a reduced-order set of nonlinear equations that captures the nonlinear dynamics of wave groups and most critically their transition from defocusing to focusing. This reduced-order model which represents information for the dynamics of the wave groups is combined with a probabilistic analysis of the possible wave groups that can form stochastically for a given wave spectrum~\citep{cousins16}. Note that stochasticity is inevitably introduced due to the `mixing' between harmonics that propagate with different speeds due to dispersion. The resulted schemes provide a parsimonious and robust prediction scheme for unidirectional water waves.  

The main purpose of the present paper is to extend the framework of~\cite{cousins16} from their unidirectional context 
to multidirectional water waves. Several new challenges arise in this context that are absent in
the unidirectional case. In the following section we review these challenges, summarize our framework and state the assumptions under which this reduced-order framework is applicable.

\subsection{Summary of the framework}
We seek to approximate the future spatiotemporal maximum wave height of a measured
wave field by decomposing the field as the superposition of wave groups with simple
shapes. The evolution of the simple wave groups are precomputed and stored, so that the
prediction reduces essentially to an interpolation from an existing data set. 
This reduced-order approach can be divided into the following steps:
\begin{enumerate}[I.]
\item Evolution of elementary wave groups. 
\item Decomposition of random wave fields.
\item Prediction of amplitude growth.
\end{enumerate}

\begin{figure}
\centering
\includegraphics[width=.45\textwidth]{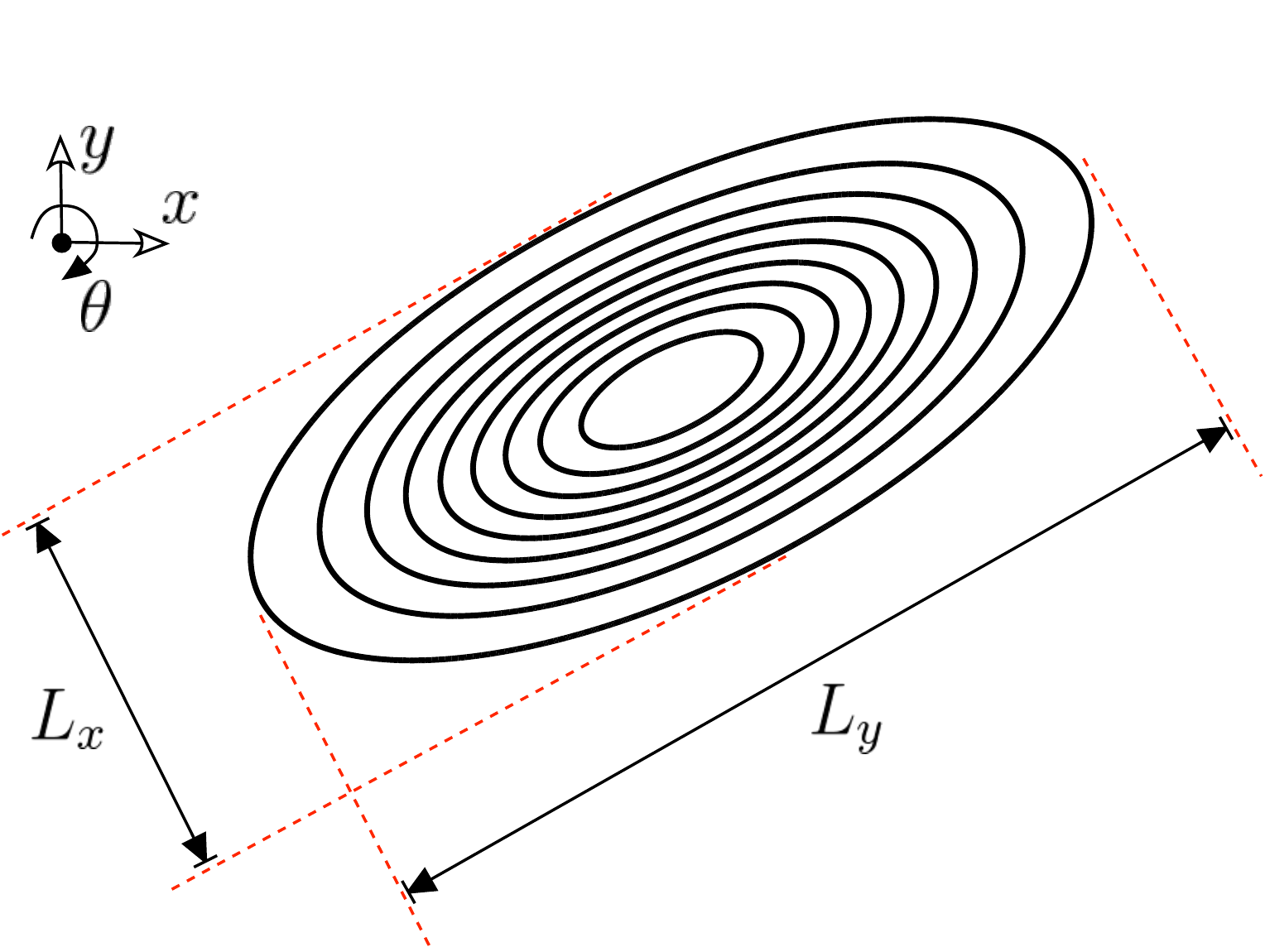}
\caption{Schematic view of an elementary wave group (EWG) with a Gaussian profile.}
\label{fig:gwg}
\end{figure}

\textbf{Step I.} We consider spatially localized simple wave groups that can be expressed analytically and refer to them as
\emph{elementary wave groups} (EWG). In this paper we will use EWG with a Gaussian profile. One can
alternatively use other shapes such as the secant hyperbolic used in~\cite{cousins16}. 
The key requirement is that the EWG must be completely
determined with only a few parameters. A Gaussian wave group, for instance, is determined by its
amplitude ($A_0$), its longitudinal and transverse widths ($L_x$ and $L_y$) and its orientation ($\theta$) with respect to
a global reference frame (see figure~\ref{fig:gwg}). Working with the Gaussian is also convenient since its derivatives with
respect to parameters and variables take a simple form.

For a realistic range of these parameters, we evolve the corresponding elementary wave groups for $T$ time units
by numerically solving an appropriate wave envelope equation (see Section~\ref{sec:env}). We record the spatiotemporal maximum amplitude 
$\Amax$ that each EWG reaches over the time interval $[0,T]$. This step is computationally expensive but is carried out 
only once. The resulting maximal amplitude is stored as a function of the parameters, i.e., $\Amax(A_0,L_x,L_y,\theta)$.
This step is carried out in Section~\ref{sec:ewg}.

\textbf{Step II.} Given a measured wave field, we approximate its envelope 
as a superposition of the elementary wave groups.
To this end, the initially unknown positions, amplitudes and length scales of the elementary wave 
groups need to be determined such that their superposition provides 
a reasonable approximation of the measured wave field. In Section~\ref{sec:decomp}, we devise a dynamical 
systems-based method that determines the unknown parameters at a reasonable computational cost.

\textbf{Step III.} Once the measured wave field is decomposed into EWGs, its maximal future amplitude
can be approximated by simply evaluating the precomputed function $\Amax(A_0,L_x,L_y,\theta)$ for each set 
of EWG parameters $(A_0,L_x,L_y,\theta)$.

\ \\

The above approach implicitly assumes that rogue waves form from individual wave groups 
through modulation instability~\citep{benjamin67}; a mechanism that has been observed 
in numerical simulations and in experiments~\citep{tulin1999,chabchoub2016}.
It is known from linear random wave theory that rogue waves can also form due to the 
constructive interference of small-amplitude wave packets~\citep{longuet1952}, especially for the case of short crested waves where modulation instability is not very pronounced. The probability of a rogue wave occurring through this 
linear mechanism, however, is orders of magnitude smaller than the ones generated through modulation 
instability~\citep{onorato2004,shemer2010,xiao2013}.
We therefore neglect the rogue waves formed from superposition of smaller waves, and focus on the modulation instability 
of individual wave groups.

Also implicit in our approach is the assumption that the wave groups constructing a wave field have negligible 
interactions over the time interval $[0,T]$. This is of course not the case for large $T$. \cite{cousins16} find, however, that for intermediate 
time scales (10-40 wave periods) and for unidirectional waves this assumption is reasonable. We come to a similar 
conclusion for the two-dimensional surface waves considered here (see Section~\ref{sec:predict}).

\section{Envelope equation}\label{sec:env}
To the first order, the modulations of a wave train with characteristic wave vector $\vc k_0$ and frequency 
$\omega_0=\omega(\vc k_0)$ can be written as
$\zeta(\vc x,t)=\frac{1}{2}u(\vc x,t)\exp[i(\vc k_0\cdot\vc x-\omega_0 t)] + \mbox{c.c.},$
where $u$ is the complex valued envelope for the modulation of the wave and c.c. is shorthand for
complex conjugate terms. Assuming that the modulations are slowly varying
(compared to the carrier wave), one can derive an equation for the envelope $u$ of the
free surface height $\zeta$.

As shown by~\cite{trulsen2000}, for deep water, the most general form of the envelope equation can be written as 
\begin{equation}
u_t+\frac{1}{4\pi^2}\int i (\omega(\vc k+\vc k_0)-\omega_0)e^{i\vc k\cdot(\vc x-\vc \xi)}
u(\vc \xi,t)\,\id\vc \xi\,\id\vc k+N(u)=0,
\label{eq:generalEnvEq}
\end{equation}
where $N(u)$ denotes the nonlinear terms to be discussed shortly. The integral term is the exact form of
the linear dispersion.
The frequency $\omega(\vc k+\vc k_0)$ can be expanded in Taylor series around wavenumber $\vc k_0$. Truncating the 
series, and accordingly the nonlinear term $N(u)$, one obtains various approximate equations for the
envelope $u$.

For instance, truncating the series at the second order and keeping the simplest cubic nonlinearity, one obtains the
nonlinear Schr\" odinger (NLS) equation~\citep{zakharov68}. 
The NLS equation is only valid for wave spectra with a narrow bandwidth.
To relax this limitation,~\cite{dysthe79} derived the modified nonlinear Schr\" odinger (MNLS) equation by 
including higher-order terms in the truncation of~\eqref{eq:generalEnvEq}. 
Rotating the spatial coordinate system $\vc x=(x,y)$ such that $\vc k_0=(k_0,0)^\top$, 
and normalizing the space and time variables as $t\mapsto \omega_0 t$ and $\vc x\mapsto k_0\vc x$,  
the MNLS equation reads
\begin{align}
u_t+\frac{1}{2}u_x+\frac{i}{8}u_{xx} & -\frac{i}{4}u_{yy}-\frac{1}{16}u_{xxx}+\frac{3}{8}u_{xyy}\nonumber\\
 &+\frac{i}{2}|u|^2 u +\frac{3}{2}|u|^2u_x+\frac{1}{4}u^2u_x^\ast + iu\overline{\phi}_x|_{z=0}=0,
\label{eq:mnls_2d}
\end{align}
where the last term involving the velocity potential is defined,
using the Fourier transform $\mathcal F$, as
$$\overline{\phi}_x|_{z=0}=-\frac{1}{2}\mathcal F^{-1}\left[\frac{k_x^2}{|\vc k|}\mathcal F(|u|^2) \right],$$
with $\vc k=(k_x,k_y)$ being the wave vector. 
\begin{figure}
\centering
\includegraphics[width=.8\textwidth]{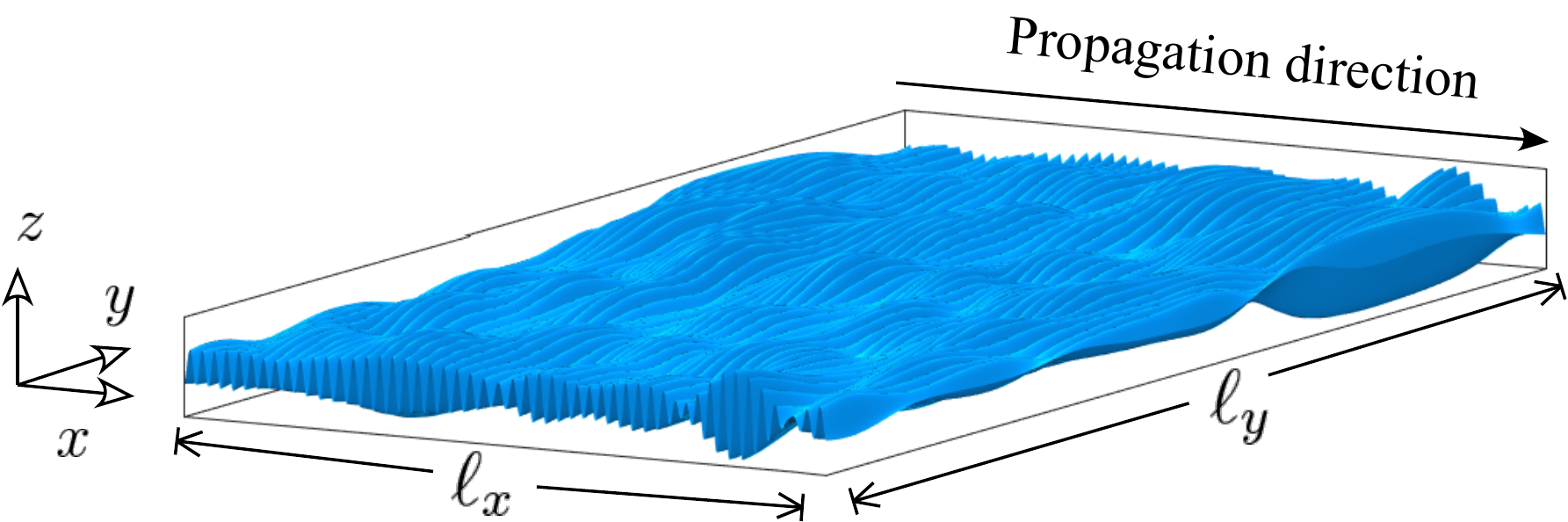}
\caption{Schematic view of the computational domain.}
\label{fig:domain}
\end{figure}

The narrow bandwidth constraint can be further improved by including even
more linear and nonlinear terms from equation~\eqref{eq:generalEnvEq}
to obtain the broad-band MNLS (BMNLS) equation~\citep{trulsen96}.
For the time scales considered here, however, the MNLS equation~\eqref{eq:mnls_2d} provides an adequate model of 
gravity waves in deep ocean~\citep{trulsen01,xiaoThesis}.

We numerically integrate the MNLS equation~\eqref{eq:mnls_2d} on a finite two-dimensional domain
$(x,y)\in[0,\ell_x]\times [0,\ell_y]$ with periodic boundary conditions (see figure~\ref{fig:domain}). We set
the domain size $2\ell_x=\ell_y=200\pi$ ($100$ characteristic wavelength). As is discussed in Section~\ref{sec:ewg},
wave groups which are elongated transverse to the propagation direction have a better chance to give rise to
rogue waves. The rectangular computational domain with a larger transverse dimension ($\ell_y=2\ell_x$) is considered here 
in order to allow for several of these wave packets to fit in the domain.

The right-hand side of the MNLS equation is evaluated using a standard pseudo-spectral method, where
the derivatives are computed in the Fourier domain and the nonlinear terms are computed in the 
physical domain. The temporal integration of the equations are carried out by a fourth-order
Runge--Kutta exponential time differencing (ETD4RK) scheme~\citep{cox02}. This method treats the
linear part of the MNLS equation exactly, and uses the Runge--Kutta scheme for the evolution
of the nonlinear terms. In the following computations, we use $2^{9}\times 2^{8}$ Fourier modes
to approximate the envelope $u(\vc x,t)$. 
The time-step size for the ETD4RK scheme is $\Delta t=0.025$.

\section{Evolution of elementary wave groups}\label{sec:ewg}
We consider elementary wave groups with the Gaussian profile
\begin{equation}
u_0(x,y)=A_0\exp\left[-\frac{x^2}{L_x^2} -\frac{y^2}{L_y^2} \right],
\label{eq:ic_gauss}
\end{equation}
where the parameters 
$L_x$ (controlling the width of the group in the longitudinal direction $x$),
$L_y$ (controlling the width of the group in the transverse direction $y$) and
$A_0$ (controlling the amplitude of the group)
determine the group completely. Note that, for simplicity, we set the orientation angle of
the group $\theta$ to zero (see figure~\ref{fig:gwg}). This is motivated by the fact that wave groups tend to align 
with the propagation direction of the underlying wave train.
\begin{figure}[h]
\centering
\includegraphics[width=.8\textwidth]{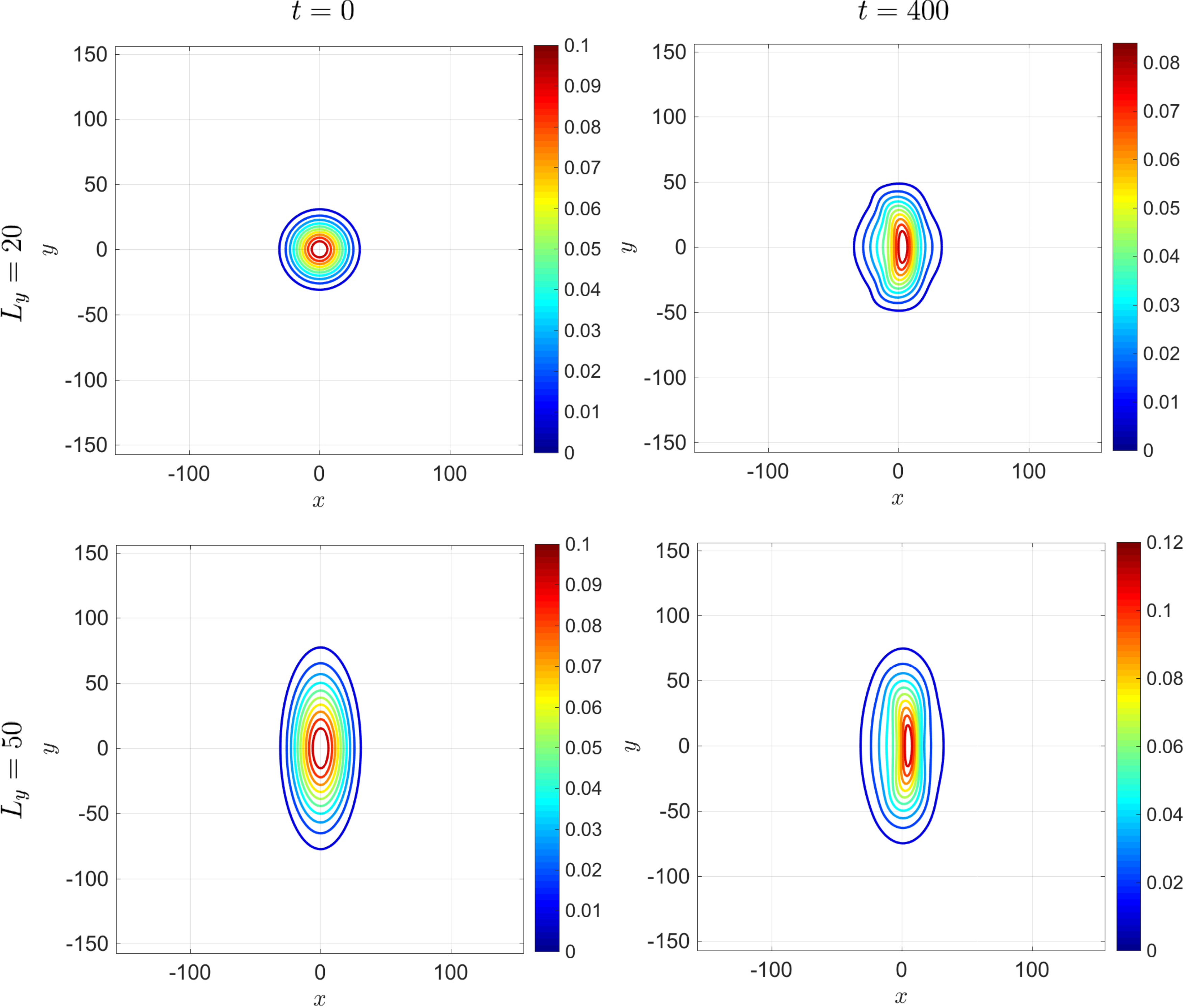}
\caption{Initial Gaussian wave group with $L_x=20$, $L_y=20$ and $A_0=0.1$ (top row)
and with $L_x=20$, $L_y=50$ and $A_0=0.1$ (bottom row).
The color marks the wave height.}
\label{fig:gewg_example}
\end{figure}

Using the MNLS equation, we evolve the initially Gaussian elementary wave groups~\eqref{eq:ic_gauss} 
for a wide range of parameters $(L_x,L_y,A_0)$. Figure~\ref{fig:gewg_example} shows 
two examples of the EWGs with the transverse 
widths $L_y=20$ and $L_y=50$. Both groups have the same longitudinal width $L_x=20$ and amplitude $A_0=0.1$.
Over time, the amplitude of the wave group with $L_y=20$ decays monotonically
and its lateral width grows slightly. The broader wave group with $L_y=50$, however, 
undergoes focusing, whereby its amplitude increases and its longitudinal width decreases slightly. At
later times $t>450$, the amplitude of this wave group decays eventually.

If instead of the initial amplitude $A_0=0.1$, we choose a smaller amplitude (say $A_0=0.05$), the 
Gaussian EWG with $L_x=20$ and $L_y=50$ would not undergo amplitude growth either. These observations
indicate that the focusing (or defocusing) of Gaussian EWGs depends non-trivially on 
all three parameters $(L_x,L_y,A_0)$. \cite{cousins16} observe the same phenomena 
in the case of unidirectional waves, although in that case the parameter $L_y$ is absent.

In order to analyze this parametric dependence systematically, we evolve the wave groups~\eqref{eq:ic_gauss}
for a range of parameters $(L_x,L_y,A_0)$. The integration time here is $T=1500$ (approximately $240$ wave periods). 
For each parameter set, we record the maximum 
wave amplitude attained by the wave group,
\begin{equation}
\Amax := \max_{x,y,t}|u(x,y,t)|,
\label{eq:Amax}
\end{equation}
where the maximum is taken over $x\in [0,\ell_x]$, $y\in[0,\ell_y]$ and $t\in[0,T]$.
Note that the maximal amplitude $\Amax$ is a function of the parameters $(L_x,L_y,A_0)$.
We also define the amplitude ratio $r=\Amax/A_0$ between the maximal amplitude $\Amax$ 
and the initial amplitude $A_0$. It is clear from definition~\eqref{eq:Amax}
that $r\geq 1$. The values $r>1$ indicate a focusing wave group, that is the amplitude of the EWG has
increased over the time interval $[0,T]$. 

\begin{figure}
\centering
\subfloat[]{\includegraphics[width=.45\textwidth]{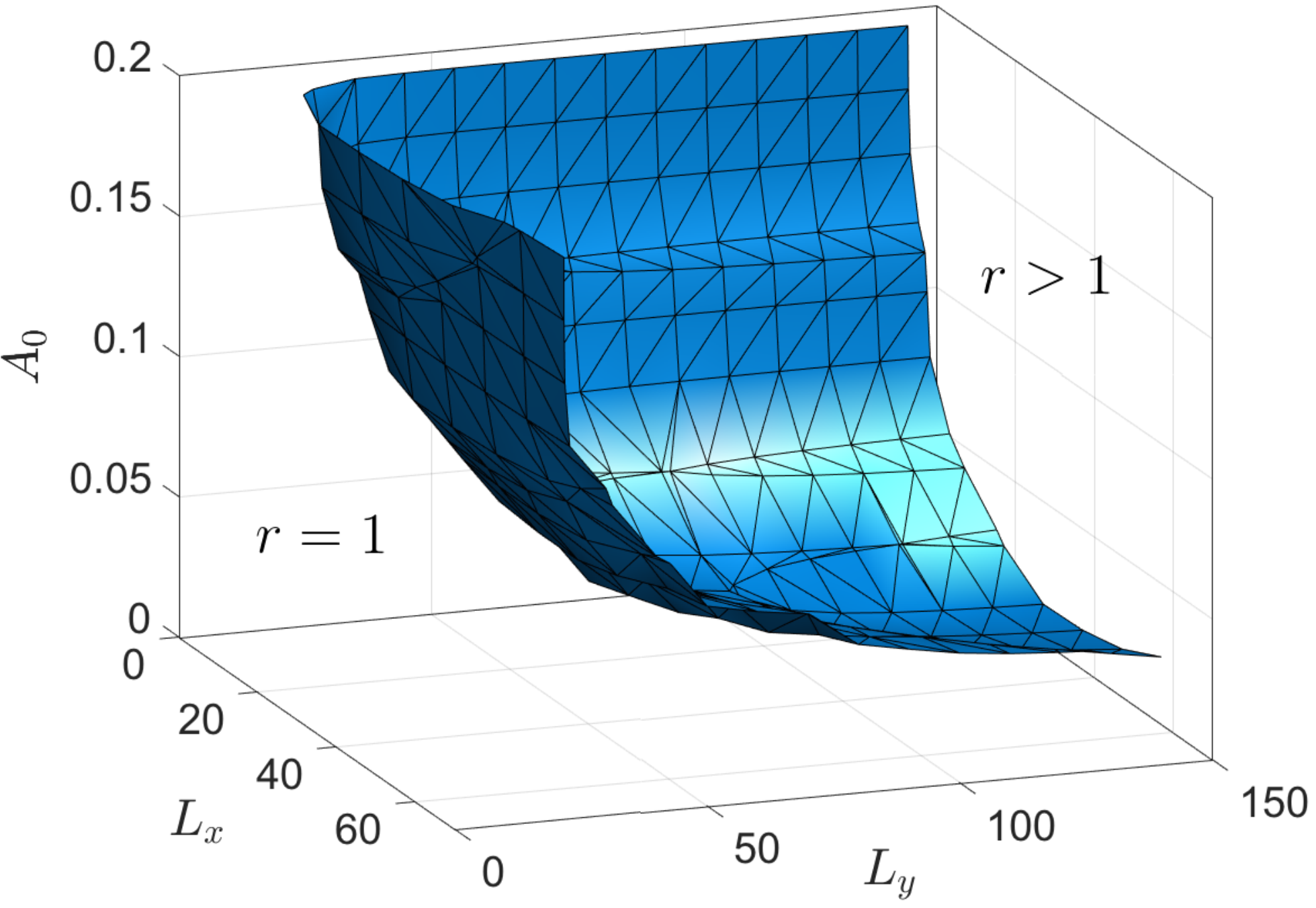}}
\subfloat[]{\includegraphics[width=.5\textwidth]{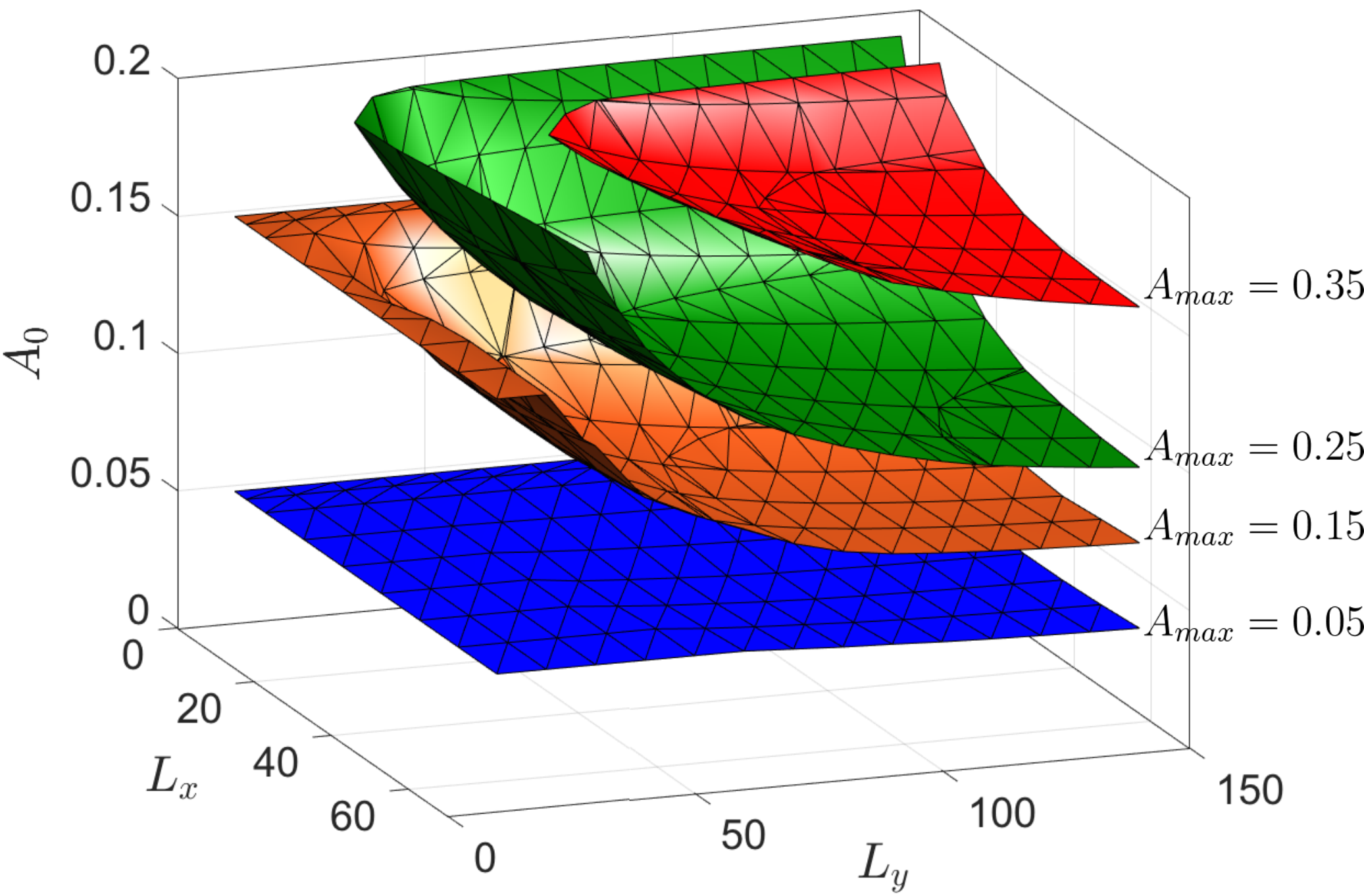}}
\caption{
(a) The hypersurface dividing the region between no ampliude growth ($r=1$) from the region with
amplitude growth ($r>1$), where $r$ is the amplitude ratio $r=\Amax/A_0$.
(b) Iso-surfaces of $A_{max}=\max_{x,y,t}|u(x,y,t)|$ corresponding to 
$A_{max} = 0.05$ (blue), 
$A_{max} = 0.15$ (orange),
$A_{max} = 0.25$ (green) and
$A_{max} = 0.35$ (red)
}
\label{fig:instBoundary}
\end{figure}

Figure~\ref{fig:instBoundary}(a) shows the hypersurface that forms the boundary between $r=1$ and $r>1$
as a function of the parameters $(L_x,L_y,A_0)$. The elementary wave groups corresponding to the points
below this surface do not undergo amplitude growth, while the points above the surface do. 
In other words, the displayed hypersurface forms the instability boundary for the EWGs in the parameter space
$(L_x,L_y,A_0)$.

Figure~\ref{fig:instBoundary}(a) shows that the amplitudes of the EWGs with very small width in $x$ or in $y$ 
do not increase. More precisely, if $L_x<2$ or $L_y<10$, the EWG's amplitude decays, irrespective of the 
initial amplitude $A_0$. As it has been shown for unidirectional waves \citep{cousins15}, this is a direct consequence of the scale-invariance breaking due to the additional terms of MNLS (compared with NLS). Here we observe the corresponding result for two-dimensional waves. For larger values of $L_x$ and $L_y$,  the amplitude growth (or lack thereof)
depends on the initial amplitude. If the initial amplitude is too small (i.e., $A_0\leq 0.05$), the wave amplitude will not 
grow at later times, irrespective of $L_x$ and $L_y$. But if the initial amplitude exceeds a threshold (depending on $L_x$
and $L_y$) we observe an amplitude growth.
\begin{figure}
\centering
\includegraphics[width=.5\textwidth]{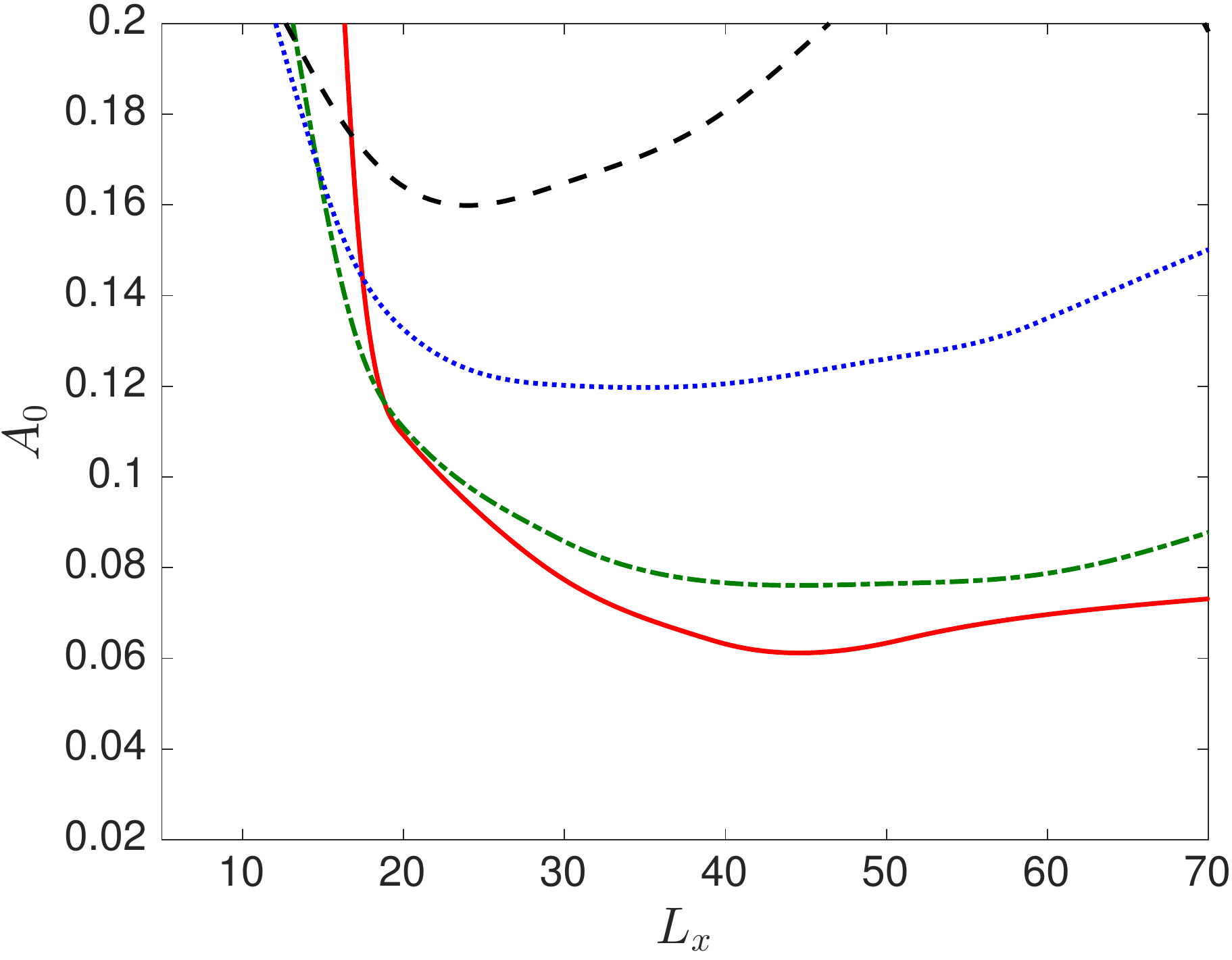}
\caption{The instability boundaries of Gaussian wave groups for various $L_y$. Each curve corresponds to a 
$L_y=\mbox{constant}$ slice of figure~\ref{fig:instBoundary}(a). 
Below each curve we have $r=1$ (no focusing) while above the curve we have $r>1$ (focusing). 
The curves correspond to
$L_y=15$ (dashed, black), 
$L_y=30$ (dotted, blue),
$L_y=60$ (dash-dotted, green) and
$L_y=120$ (solid, red).
}
\label{fig:instBoundaryLxA0}
\end{figure}

This is further demonstrated in figure~\ref{fig:instBoundary}(b), showing a few iso-surfaces of the 
maximum amplitude $\Amax$ as a function of the parameters $(L_x,L_y,A_0)$. We observe that the 
iso-suface $\Amax=0.05$ coincides with the surface $A_0=0.05$. This indicates that, regardless of the
initial withs $L_x$ and $L_y$, the amplitudes of the EWGs decay if the initial amplitude is small enough.

Iso-surfaces corresponding to larger values of $\Amax$ exhibit curved manifolds that correspond to
the required combination of parameters $(L_x,L_y,A_0)$ for the Gaussian EWG to reach the 
amplitude $\Amax$ at some time in the interval $[0,T]$.

For a finite $L_x$ as the width $L_y$ tends to infinity, we approach the unidirectional waves. 
Figure~\ref{fig:instBoundaryLxA0} shows the instability boundary as 
one approaches this unidirectional limit. The solid curve (red color) corresponding
to $L_y=120$ is in agreement with available unidirectional results 
(cf. figure 2(b) from~\cite{cousins16} and figure 3 of~\cite{cousins15}).

In Section~\ref{sec:predict}, we use the computed maximal amplitude $\Amax(L_x,L_y,A_0)$ to estimate the
amplitude growth of a given random wave field. To this end, we first need to approximate the random
wave fields as a superposition of Gaussian EWGs (see section~\ref{sec:decomp} below).
\begin{figure}
\centering
\subfloat[]{\includegraphics[width=.45\textwidth]{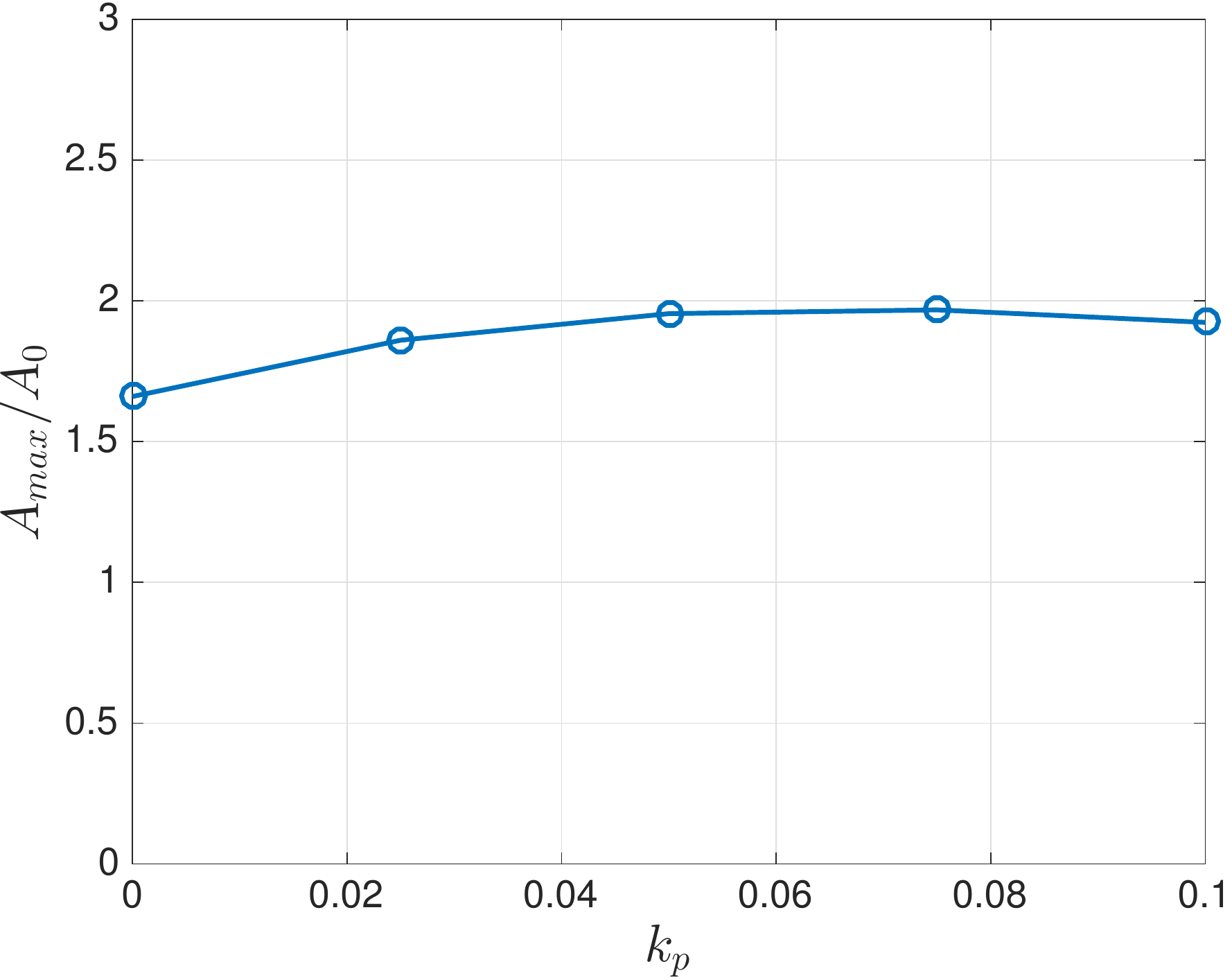}}\hspace{.05\textwidth}
\subfloat[]{\includegraphics[width=.45\textwidth]{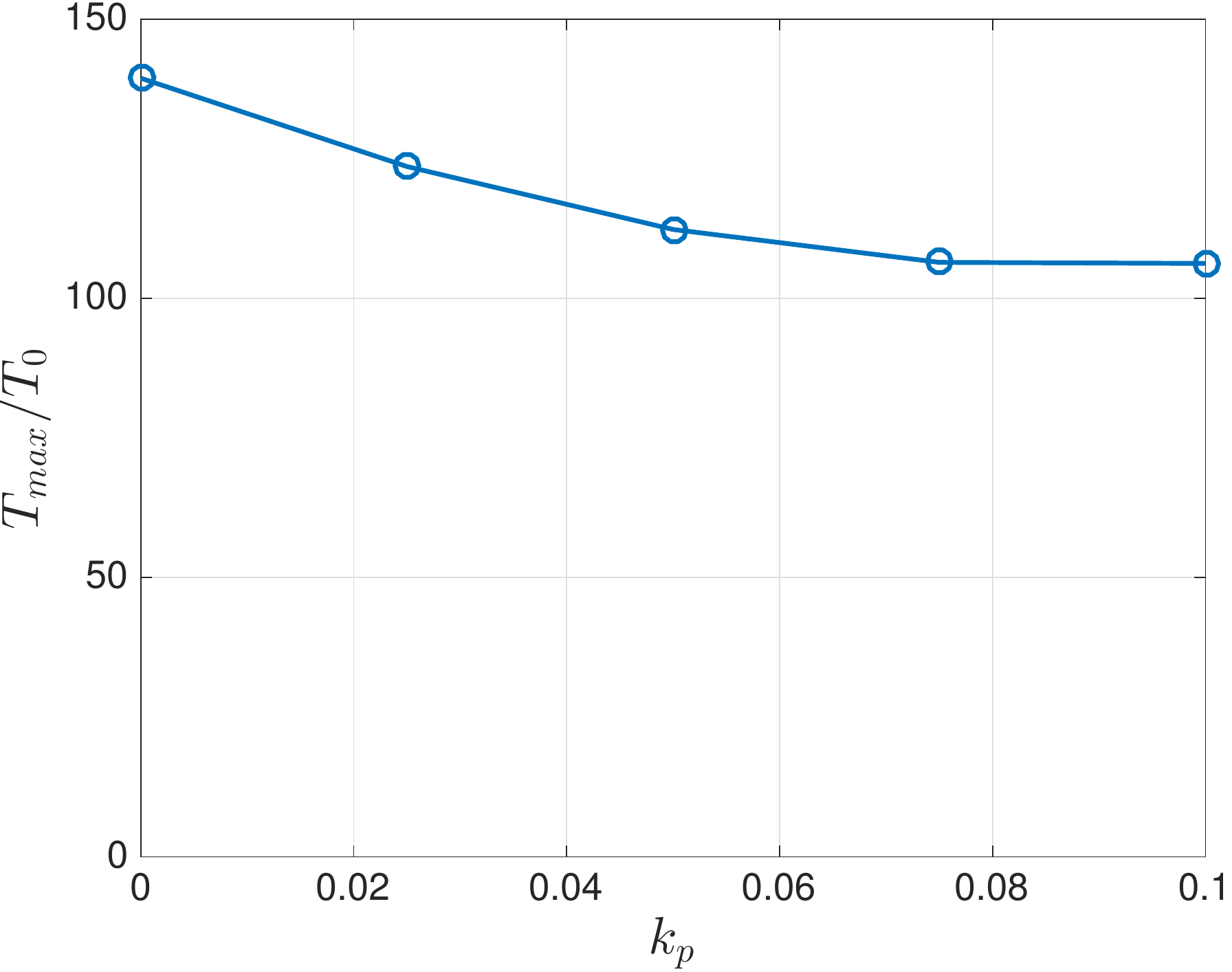}}
\caption{The evolution of phase-dependent EWG $u_0(x,y)=A_0\exp(-x^2/L_x^2)\exp(-y^2/L_y^2)\exp(i\pi k_p x)$
with $A_0=0.09$, $L_x=30$, $L_y=75$ and $0\leq k_p\leq 0.1$. 
(a) The maximum amplitude $A_{max}$ over the time interval $[0,T]$ with $T=1500$.
(b) The time $T_{max}$ when the maximum amplitude $A_{max}$ is reached.
Here, $T_0=2\pi$ is the wave period.}
\label{fig:phase}
\end{figure}

We point out that one can further generalize the choice of the EWG~\eqref{eq:ic_gauss} by introducing a phase parameter, e.g.
$u_0(x,y)=A_0\exp[-x^2/L_x^2-y^2/L_y^2]\exp[i\pi k_p x]$ where $k_p\ll k_0$ determines the
phase of the modulating wave envelope. The elementary wave group~\eqref{eq:ic_gauss} corresponds to $k_p=0$
which simplifies the decomposition of the random wave fields into EWGs by reducing the number of free parameters.
The effect of this choice ($k_p=0$) on the resulting maximal amplitude $A_{max}$ is insignificant (see figure~\ref{fig:phase}(a)). 
However, the time when this amplitude is reached depends significantly on 
the phase (see figure~\ref{fig:phase}(b)). As a result, our reduced-order prediction, which ignores the phase dependence,
predicts the occurrence of a rogue wave over the future time window $[0,T]$ but not the 
exact time of its occurrence. 

\section{Decomposition of random wave fields}\label{sec:decomp}
To apply the results of the wave group evolution we need to design a robust algorithm for the decomposition of any arbitrary wave field. Given such measured wave field, we 
approximate its envelope $|u_0|$ as a superposition of $N$ Gaussian functions,
\begin{equation}
G(x,y)=\sum_{n=1}^{N}g_n(x,y),
\label{eq:gauss_series}
\end{equation}
where $g_n$ is the $n$-th Gaussian elementary wave group,
\begin{equation}
g_n(x,y)=a_n\exp\left[-\frac{(x-x^c_n)^2}{(L^x_n)^2}\right]\exp\left[-\frac{(y-y^c_n)^2}{(L^y_n)^2}\right].
\end{equation}
The functions $g_n$ are identical to the
Gaussian EWGs~\eqref{eq:ic_gauss}, modulo a shift in the location of their peaks $(x^c_n,y^c_n)$.
The unknown parameters to be determined are
\begin{align*}
\vc a = (a_1,\cdots, a_N),\quad & \\
\vc x^c = (x^c_1,\cdots,x^c_N),\quad & \vc y^c=(y^c_1,\cdots, y^c_N),\\
\vc L^x = (L^x_1,\cdots,L^x_N),\quad & \vc L^y=(L^y_1,\cdots, L^y_N).
\end{align*}

We first determine the location
$(x^c_n,y^c_n)$ of each Gaussian group from the local maxima of the 
envelope $|u_0|$. The local maxima can be readily located 
by a peak detection algorithm as detailed in Section~\ref{app:peakDetect}.
The corresponding amplitude $a_n$ of each 
Gaussian wave group is determined by the amplitude of the envelope at
the point $(x^c_n,y^c_n)$, i.e., $a_n=|u_0(x^c_n,y^c_n)|$.

Once the centers $(\vc x^c,\vc y^c)$, and hence the amplitudes $\vc a$, are found, it remains to determine
the width parameters $(\vc L^x,\vc L^y)$ of the Gaussian profiles. To this end, we use the following optimization step.
Given the envelope $|u_0|$, we define the function
\begin{equation}
J(\vc L^x,\vc L^y):=\frac{1}{2}\int_{0}^{\ell_y}\int_{0}^{\ell_x} \big(G(x,y)-|u_0(x,y)|\big)^2\id x \id y,
\label{eq:J}
\end{equation}
where $G$ is the superposition of Gaussian wave groups~\eqref{eq:gauss_series} to be determined.
The global minimizer of the function $J:\mathbb R^N\times\mathbb R^N\to \mathbb R$ returns the best Gaussian approximation to the envelope $|u_0|$. 

Minimizing the functional $J$ is a standard optimization problem.
Here, we evaluate the minimizers by devising an appropriate 
fictitious-time differential equation as described in Section~\ref{app:lengthDetect}.
This differential equation is the continuous limit of the gradient descent method
and its trajectories are guaranteed to converge to stationary points of the functional $J$ (see, e.g., section 6-6b of~\cite{pierre1969} or \cite{faraz_adjoint}). Before discussing the optimization method, however, we need to locate the peaks $(x_n^c,y_n^c)$.

\subsection{Detection of the peaks}\label{app:peakDetect}
There are several methods for detection of local maxima of a two-dimensional surface~\citep{recipe}.
Here, we approximate the local maxima of the envelope $|u_0|$ by simply comparing the nearest neighbors on the given
computational grid. On a rectangular grid, we declare a grid point a local maximum if the value of the envelope 
is larger than its eight immediate neighbors. For our purposes this rudimentary method returns satisfactory results 
and avoids the computational cost of more high-end peak detection algorithms.

Before applying this peak detection algorithm, however, we apply a low-pass filter to the measured envelope $|u_0|$.
This filter is not an ad hoc smoothing; it is rather motivated by the observations made in Section~\ref{sec:ewg}. Recall that
wave groups with lengths scales $L_x<2$ or $L_y<10$ decay regardless of their initial amplitude $A_0$. Since our purpose 
is to predict the growth of wave groups, we can safely neglect such small scale wave groups. Given this observation, therefore, 
we discard harmonics whose wavenumbers $(k_x,k_y)$ satisfy
\begin{equation}
\frac{1}{2}\frac{\ell_x}{2\pi}< k_x\quad \mbox{or}\quad \frac{1}{10}\frac{\ell_y}{2\pi}< k_y.
\label{eq:lowPassFilter}
\end{equation}
Note that, given the domain size $\ell_x\times \ell_y$, these wavenumbers correspond
to decaying wave groups with $L_x<2$ or $L_y<10$.

This physically-motivated smoothing has two advantages. First, it speeds up the
computations by removing many irrelevant, small-scale peaks from the envelope. Second, as we further discuss in 
Section~\ref{sec:noise}, it makes our group detection method robust to measurement noise. 

\subsection{Detection of the length scales}\label{app:lengthDetect}
Once the peaks are detected, it remains to find the optimal set of length scales $(\vc L^x,\vc L^y)$
that minimizes the function $J:\mathbb R^N\times \mathbb R^N\to \mathbb R$ defined in~\eqref{eq:J}.
Here, we devise a dynamical systems-based method that can efficiently find these minimizers.
The idea is to evolve $\vc L^x(\tau)$ and $\vc L^y(\tau)$ along the fictitious time $\tau$ such that
the functional $J(\vc L^x(\tau),\vc L^y(\tau))$ decreases monotonically as $\tau$ increases.
Differentiating with respect to the fictitious time $\tau$, we obtain
\begin{equation}
\frac{\id J}{\id \tau}=\sum_{m=1}^{N}\iint\Big(G(x,y)-|u_0(x,y)|\Big)\left(\frac{\partial G}{\partial L^x_m} \frac{\id L^x_m}{\id \tau}
+\frac{\partial G}{\partial L^y_m}\frac{\id L^y_m}{\id \tau}
\right)\id x \id y,
\end{equation}
where 
\begin{subequations}
\begin{equation}
\frac{\partial G}{\partial L^x_m} = 
2a_m\frac{(x-x^c_m)^2}{(L^x_m)^3}e^{-(x-x^c_m)^2/(L^x_m)^2}e^{-(y-y^c_m)^2/(L^y_m)^2},
\end{equation}
\begin{equation}
\frac{\partial G}{\partial L^y_m} = 
2a_m\frac{(y-y^c_m)^2}{(L^y_m)^3}e^{-(x-x^c_m)^2/(L^x_m)^2}e^{-(y-y^c_m)^2/(L^y_m)^2}.
\end{equation}
\end{subequations}
\begin{figure}
\centering
\includegraphics[width=.35\textwidth]{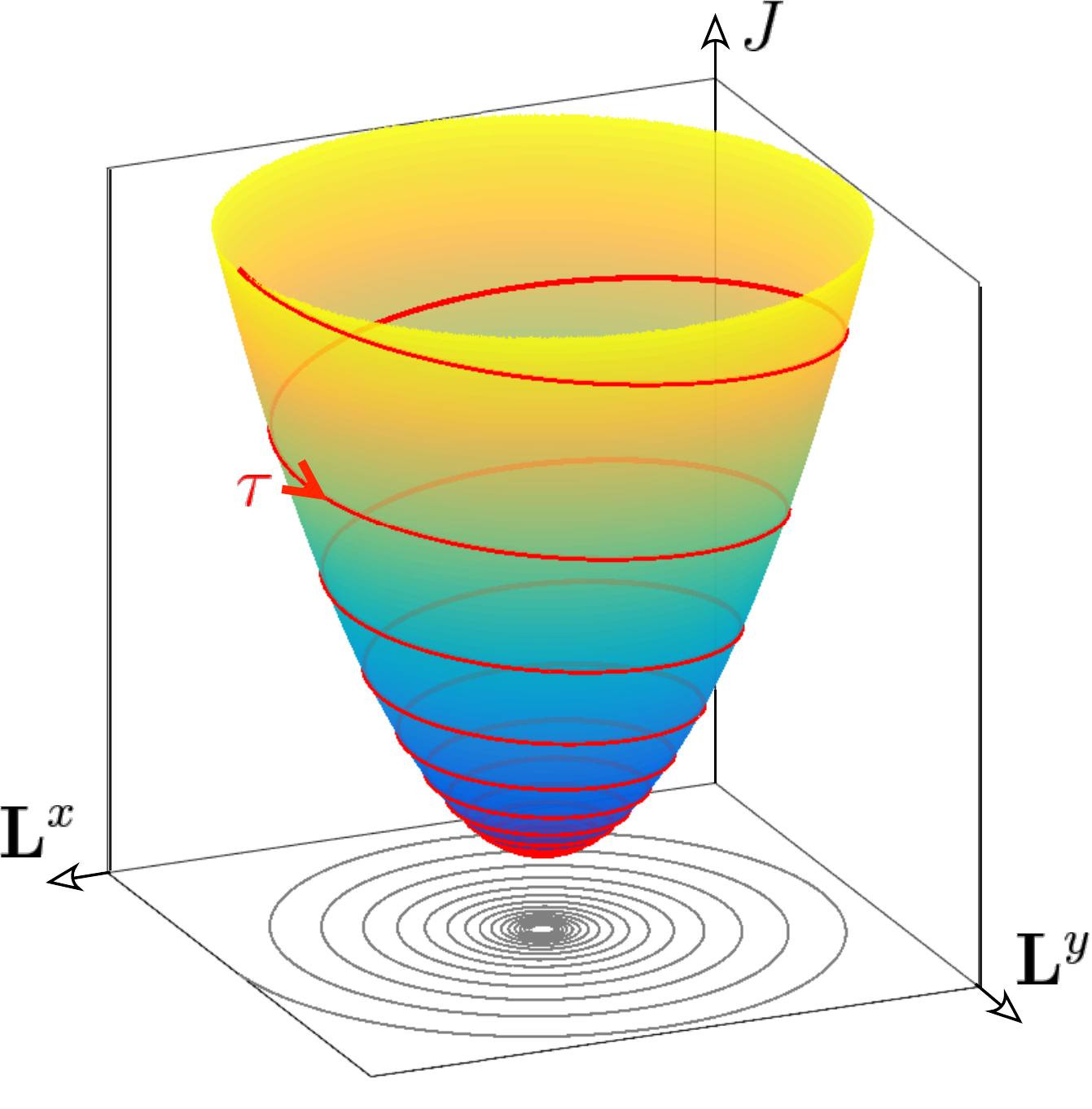}
\caption{Schematic geometry of the function~\eqref{eq:J} (color surface) and a 
trajectory of the ODE~\eqref{eq:Lode} (red curve). Note that $\vc L^x,\vc L^y\in\mathbb R^N$
and that $J:\mathbb R^{2N}\to \mathbb R$ is a multivariable function.}
\label{fig:Lode}
\end{figure}

We choose the fictitious-time derivatives $\id L^x_m/\id \tau$ and $\id L^y_m/\id \tau$
such that $J(\tau):=J(\vc L^x(\tau),\vc L^y(\tau))$ is monotonically decreasing. A trivial choice achieving this goal is
\begin{subequations}
\begin{equation}
\frac{\id L^x_m}{\id \tau}=  -\iint \Big(G(x,y)-|u_0(x,y)| \Big)\frac{\partial G}{\partial L^x_m}\id x \id y,
\end{equation}
\begin{equation}
\frac{\id L^y_m}{\id \tau}=  -\iint \Big(G(x,y)-|u_0(x,y)| \Big)\frac{\partial G}{\partial L^y_m}\id x \id y,
\end{equation}
\label{eq:Lode}
\end{subequations}
for $m=1,2,\cdots, N$.

The schematic figure~\ref{fig:Lode} shows the geometry of the function $J$ and a trajectory of the
ODE~\eqref{eq:Lode}. In reality, the graph of $J$ is much more complex with several local minima
as opposed to one global minimum depicted here. If the ODE is solved from an initial condition $(\vc L_x(0),\vc L_y(0))$
far from the global minimum, the trajectory will most likely converge to a local minimum of the function $J$ 
which could potentially result in an unsatisfactory approximation of the wave envelope. 
Therefore, it is important that the 
initial conditions $(\vc L_x(0),\vc L_y(0))$ are chosen carefully. Here, we choose these initial conditions
such that the second-order partial derivative of the Gaussian wave group $g_n(x,y)$ evaluated at the corresponding peak 
$(x_n^c,y_n^c)$ coincides with the second-order partial derivative of the measured envelope $|u_0|$ at that peak. More 
precisely, we choose $(L_n^x(0),L_n^y(0))$ such that 
\begin{subequations}
\begin{equation}
\partial_x^2g_n(x_n^c,y_n^c)= \partial_x^2|u_0|(x_n^c,y_n^c),
\end{equation}
\begin{equation}
\partial_y^2g_n(x_n^c,y_n^c)= \partial_y^2|u_0|(x_n^c,y_n^c). 
\end{equation}
\label{eq:L0}
\end{subequations}
The derivatives of the measured envelope $u_0$ are approximated 
numerically by finite differences while the derivative of $g_n$
are known analytically. 
Note that the value of the Gaussian $g_n$ at the peak $(x_n^c,y^c_n)$ is independent of the 
length scales $(L^x_n,L^y_n)$. Similarly, the first-order partial derivatives of $g_n$ vanish 
at the peaks and therefore are independent of the length scales $(L^x_n,L^y_n)$.
The lowest-order derivatives that depend on the length scales are the second-order derivatives. 
That is our motivation for using these derivatives to obtain good initial conditions $(L_n^x(0),L_n^y(0))$.

This choice of the initial conditions results in a reasonable approximation of the wave envelope $|u_0|$ such that
all the envelopes reported in Section~\ref{sec:predict} below are reconstructed with relative error,
\begin{equation}
e_r = \frac{J}{\frac{1}{2}\iint|u_0(x,y)|^2\id x \id y},
\label{eq:er}
\end{equation}
smaller than $10\%$. 

In order to evolve the ODE~\eqref{eq:Lode}, we use adaptive time stepping where the time step
$\delta\tau$ is adjusted adaptively to ensure that the relative error $e_r$ decreases after each time step.
We elaborate this adaptive time stepping in Algorithm~\ref{alg:groupDetect} 
where our wave group detection is summarized.
\begin{algorithm}[t]
\DontPrintSemicolon
\caption{Wave group detection algorithm. The variables $e_r^{(1)}$ and $e_r^{(2)}$
are relative errors which are evaluated according to equation~\eqref{eq:er}.
The inputs are the measured wave envelope $|u_0|$,
an initial time step size $\delta\tau_0$, tolerance $\texttt{Tol}$ 
and the maximum number of interations $\texttt{MaxIter}$. 
$\texttt{ODEstep}(\vc L^x(0),\vc L^y(0),\delta\tau)$ denotes an explicit ODE time-stepping schem 
for~\eqref{eq:Lode} with initial conditions $(\vc L^x(0),\vc L^y(0))$ and time step size $\delta\tau$. 
Here, the computations are carried out with 
$\delta\tau_0=2$,
$\texttt{Tol}=0.1$, 
$\texttt{MaxIter}=10^5$ and the fourth-order Runge–Kutta for $\texttt{ODEstep}$.
}
\textbf{Input:} $|u_0|$, $\delta\tau_0$, \texttt{Tol}, \texttt{MaxIter}\;
Detect peaks $(\vc x^c,\vc y^c,\vc a)$ (see Section~\ref{app:peakDetect})\;
Determine $(\vc L^x(0),\vc L^y(0))$ according to~\eqref{eq:L0}\;
Evaluate $e_r^{(1)}$ with $J=J(\vc L^x(0),\vc L^y(0))$\;
$i=0$\;
\While{\textnormal{$e_r^{(1)}>\texttt{Tol}$ and $i\leq \texttt{MaxIter}$}}{
$\delta\tau \longleftarrow \delta\tau_0$\;
$e_r^{(2)}\longleftarrow e_r^{(1)}$\;
\While{\textnormal{$e_r^{(2)}\geq e_r^{(1)}$}}{
$\delta \tau\longleftarrow \delta\tau/2$\;
$(\vc L^x,\vc L^y)\longleftarrow\texttt{ODEstep}(\vc L^x(0),\vc L^y(0),\delta\tau)$\;
Evaluate $e_r^{(2)}$ with $J=J(\vc L^x,\vc L^y)$\;
}
$(\vc L^x(0),\vc L^y(0))\longleftarrow(\vc L^x,\vc L^y)$\;
$e_r^{(1)}\longleftarrow e_r^{(2)}$\;
$i\longleftarrow i+1$\;
}
\textbf{Output:} $\vc x^c$, $\vc y^c$, $\vc a$, $\vc L^x$, $\vc L^y$, $e_r^{(1)}$
\label{alg:groupDetect}
\end{algorithm}

The accurate evaluation of the integrals in equation~\eqref{eq:Lode} requires the wave envelope $u_0$ to be  
measured on a sufficiently dense spatial grid. The available wave gauges are capable of such high-resolution 
measurements~\citep{story11,borge2013}. However, if the wave measurements are only available
on a sparse staggered grid, the parameters in the series~\eqref{eq:gauss_series} need to 
be estimated by an alternative method such as the statistical techniques of model inference
(see Chapter 2 of~\cite{bishop1995} and Chapter 8 of~\cite{friedman2001} for a survey of these statistical methods).
Here, we assume that a high-resolution measurement of the wave envelope is available so that 
the right-hand sides of differential equations~\eqref{eq:Lode} can be evaluated accurately.

\subsection{Sensitivity to measurement noise}\label{sec:noise}
In practice, the initial wave envelope $u_0$ is measured through a wave gauge 
with the ability to record the spatial surface elevation (see, e.g.,~\cite{story11,fedele2011,borge2013}).
Such measurements are inevitably contaminated with some degree of noise. It is therefore important to verify whether 
our wave group decomposition is robust with respect to such measurement noise.

\begin{figure}
\centering
\subfloat[]{\includegraphics[width=.45\textwidth]{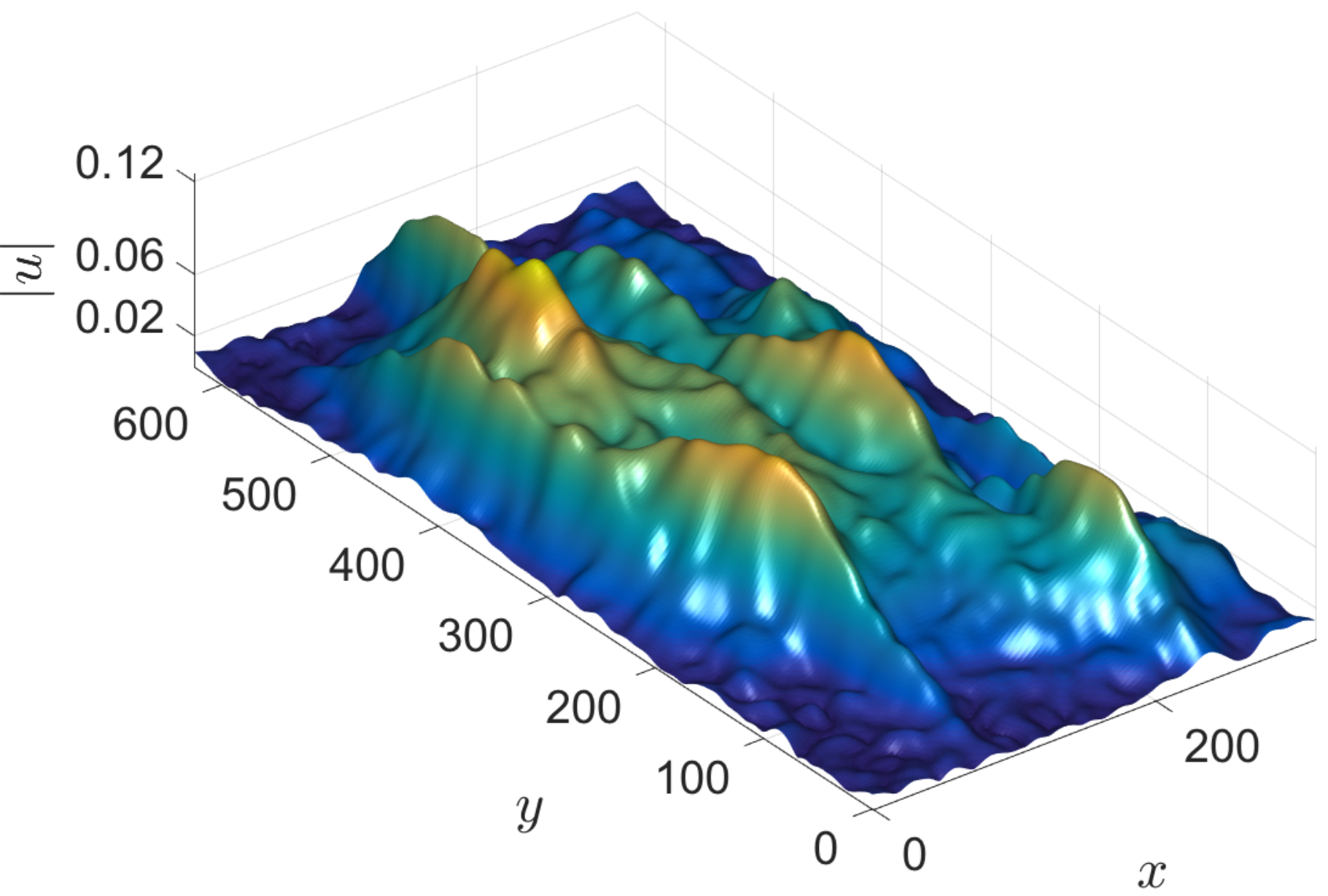}}
\subfloat[]{\includegraphics[width=.45\textwidth]{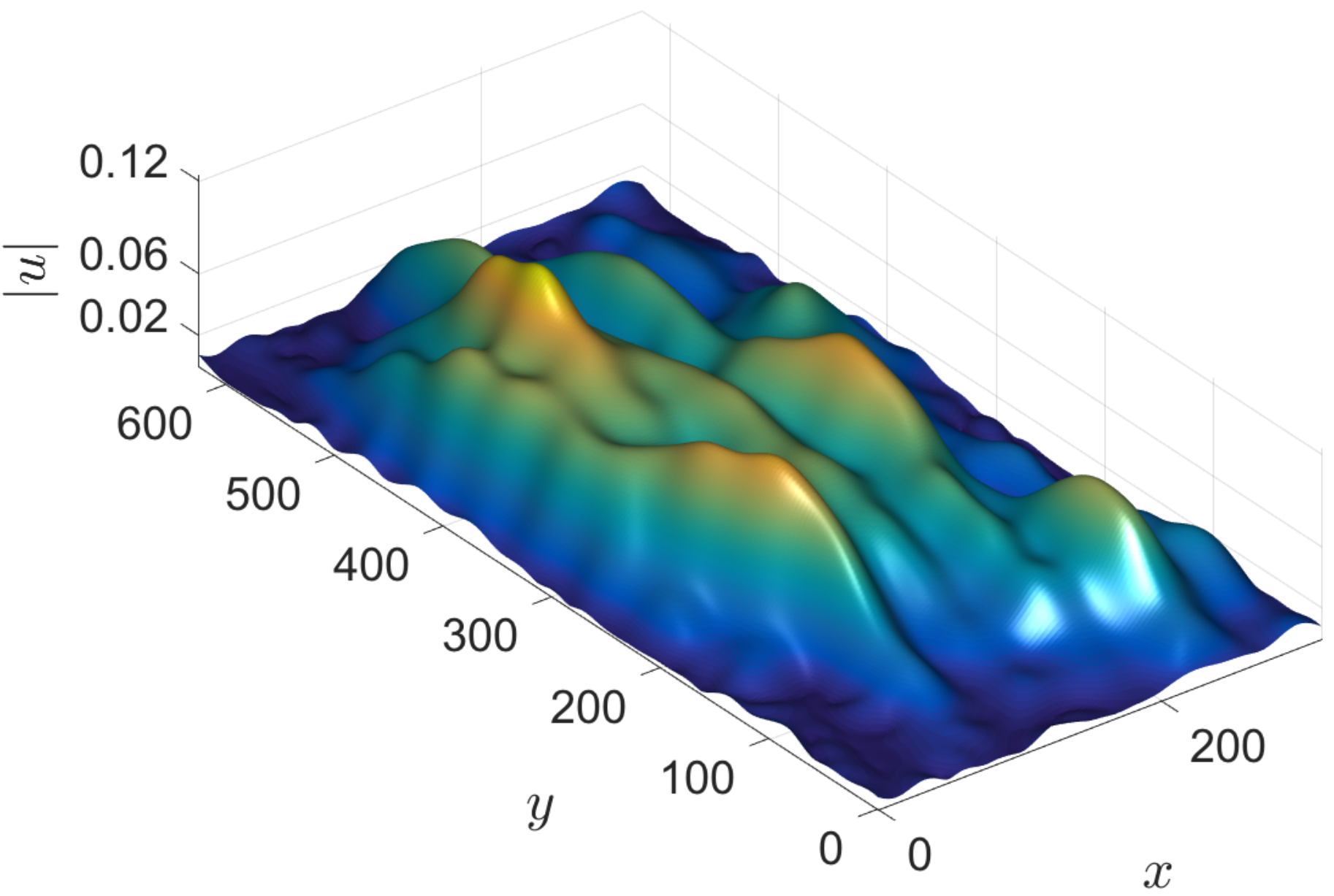}}\\
\subfloat[]{\includegraphics[width=.45\textwidth]{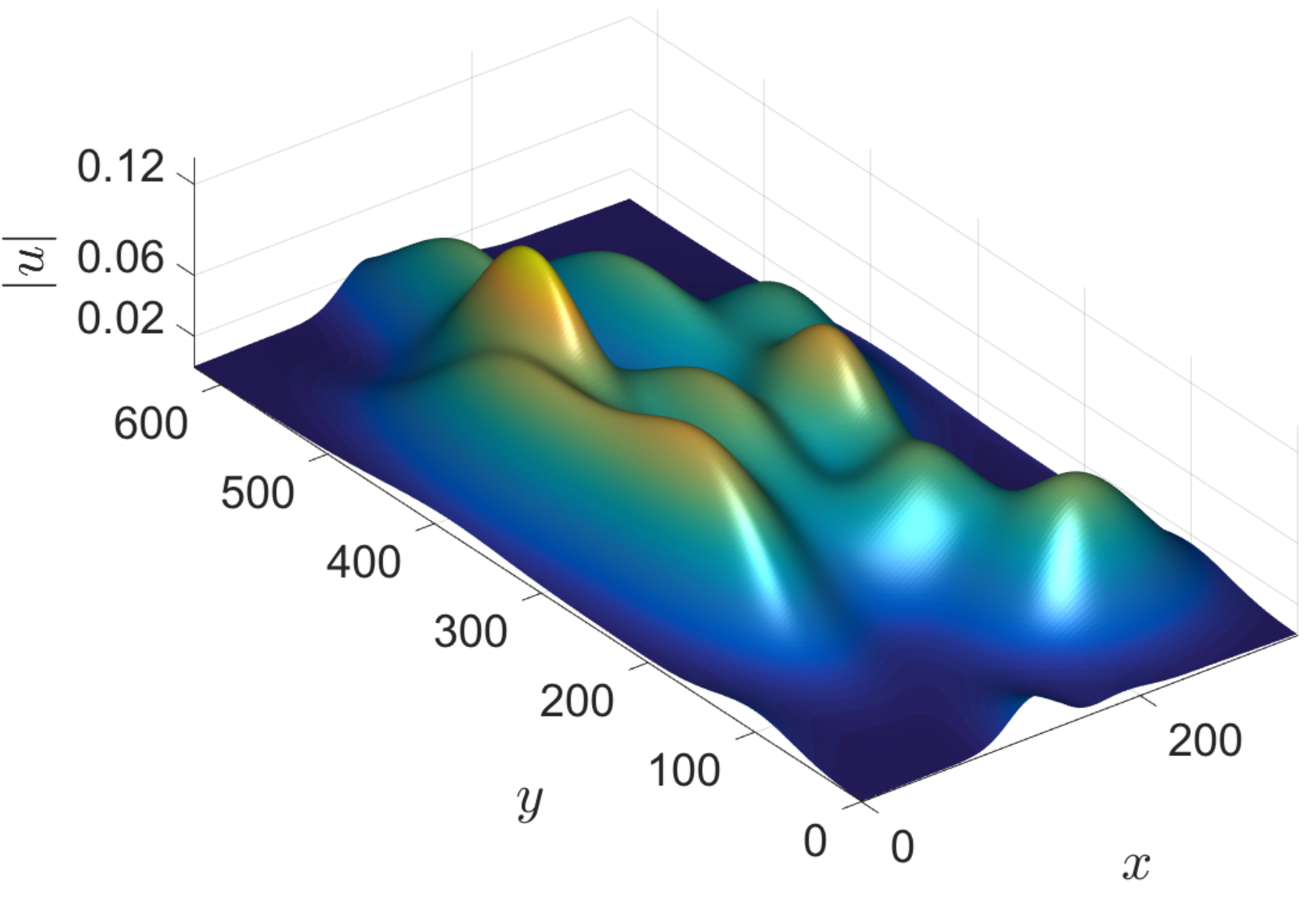}}
\subfloat[]{\includegraphics[width=.4\textwidth]{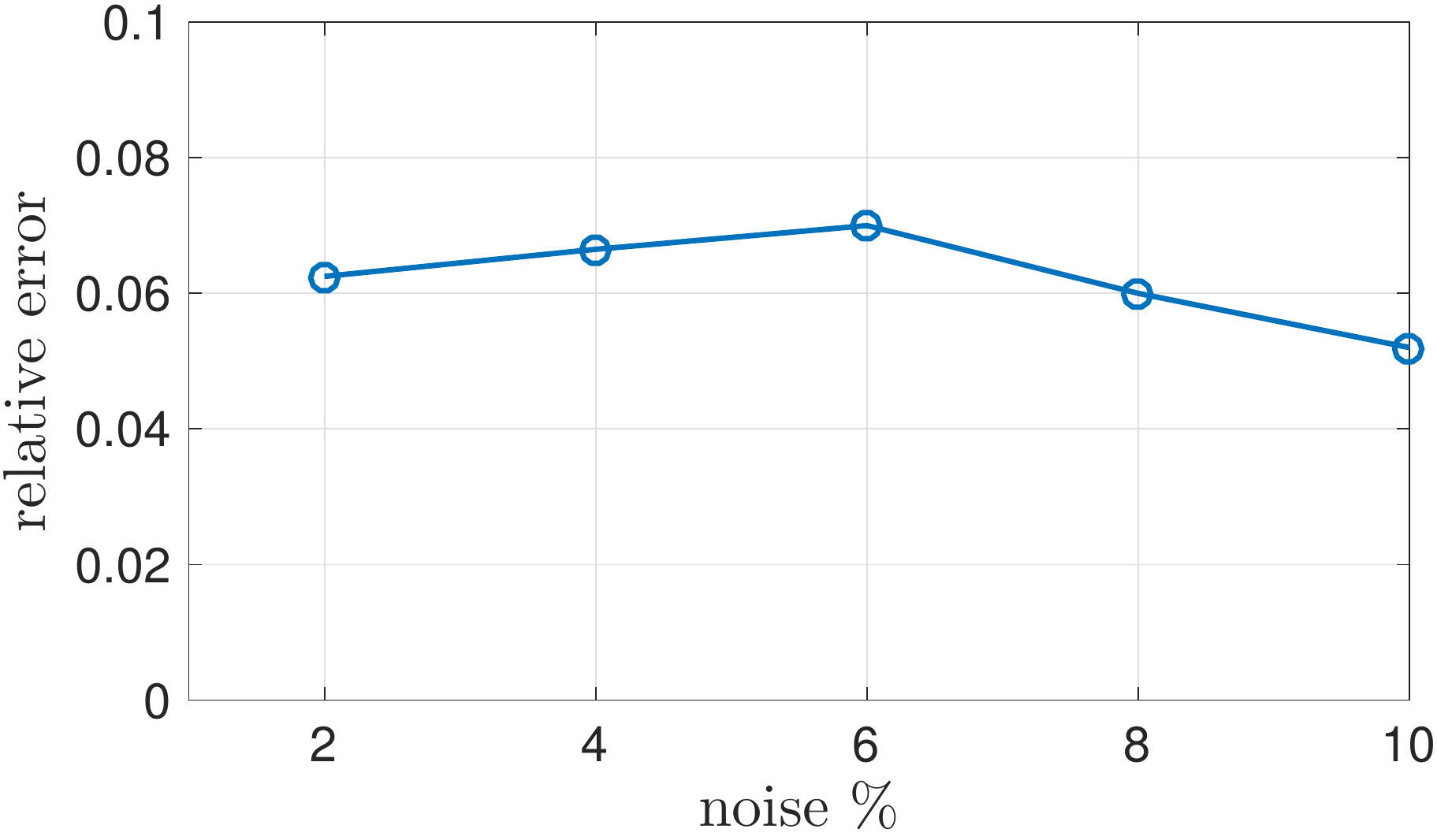}}
\caption{
(a) A wave field with $10\%$ noise added to it. 
(b) The wave field after low-pass filtering.
(c) Reconstructed wave group with Gaussian wave group approximation~\eqref{eq:gauss_series}.
(d) The relative error $e_r$ of the recostructed wave feilds as a function of the noise to signal ratio. 
}
\label{fig:groupDet_noise}
\end{figure}

To this end, we consider a noise-free envelope $u_G$ consisting of the superposition of $50$ Gaussian wave groups
whose amplitudes, locations and length scales are chosen randomly.
We add some artificial noise to this envelope to obtain the noisy envelope $u_0=u_G+u_{noise}$. The noise $u_{noise}$, 
which is correlated in space,
is obtained from a slowly decaying Gaussian spectrum with random phase. We use the ratio of the r.m.s. of $|u_{noise}|$ to
the r.m.s. of $|u_G|$ as the noise to signal ratio to quantify the strength of the noise. Figure~\ref{fig:groupDet_noise}(a)
shows an example of such a wave filed with $10\%$ noise. Panel (b) shows the wave field after applying the low pass filter
discussed in Section~\ref{app:peakDetect} 
(cf. equation~\eqref{eq:lowPassFilter}). Panel (c) shows the reconstructed wave field from the
Gaussian wave field approximation~\eqref{eq:gauss_series}. The relative error $e_r$ of 
this approximation is about $5.6\%$.

Figure~\ref{fig:groupDet_noise}(d) shows the relative error $e_r$ (see equation~\eqref{eq:er})
of the reconstructed wave fields for several levels of signal to noise ratio.
The reconstructed wave fields have relative errors lower than $8\%$ which is 
quite satisfactory given the fact that the wave gauges measure the surface elevation 
within $10\%$ error to begin with~\citep{story11}.

The robustness of our wave field decomposition can be attributed to two features of the method.
Firstly, the low pass filter discussed in section~\ref{app:peakDetect} results in a relatively smooth
surface by removing the small scale fluctuations
in the wave field (see figure~\ref{fig:groupDet_noise}(b)). 
As observed in Section~\ref{sec:ewg} (cf. figure~\ref{fig:instBoundary}), 
whether these small scale fluctuations are attributed to noise
or are genuine features of the wave field, they will not develop into rogue waves and they will not influence larger wave groups on their future evolution.
Therefore, in the context of rogue wave detection, this low pass filter is justified. 

The second reason for the robustness of our method is the spatial average taken in the cost function~\eqref{eq:J}.
As a result, the right-hand side of the ODEs~\eqref{eq:Lode} involves an integral over space with an exponential kernel.
This averaging adds an extra level of smoothing to our method. As opposed to the low pass filter, this smoothing
is not applied directly to the measured wave field $u_0$; it is instead embedded in the minimization 
method itself.

\section{Results and discussion}\label{sec:predict}
In this section, we examine the forecast skill of our method. 
To this end, we generate a large number of random wave fields that follow prescribed spectra. 
Then we decompose each field into Gaussian elementary wave groups
using the method developed in Section~\ref{sec:decomp}. The future maximal amplitude of the
random wave field is then estimated by interpolating the precomputed data $\Amax(L_x,L_y,A_0)$
from Section~\ref{sec:ewg}.

\subsection{Wave spectra}
For the wave field, we consider envelopes of the form
\begin{equation}
\widehat{u}_0(\vc k)=\frac{2\pi}{\sqrt{\ell_x\ell_y}}
\psi(\vc k)
\exp[i\phi(\vc k)],
\label{eq:env_Fc}
\end{equation}
where $\widehat{u}_0(\vc k)$ denotes the Fourier coefficient of the envelope corresponding to
the wave vector $\vc k=(k_x,k_y)\in\mathbb Z^2$.
The phase $\phi(\vc k)$ is a random variable uniformly distributed over the interval $[0,2\pi]$.
The spectrum of the waves generated from~\eqref{eq:env_Fc} coincide with $|\psi(\vc k)|^2$.

We will consider two types of wave spectra: a Gaussian spectrum and the Joint North Sea Wave Observation Project
(JONSWAP) spectrum. Following~\cite{dysthe03}, the Gaussian spectrum is defined as
\begin{equation}
\psi(\vc k)=\frac{\epsilon}{\sqrt{2\pi \sigma_x\sigma_y}}
\exp\left[-\frac{(q_xk_x)^2}{4\sigma_x^2}-\frac{(q_yk_y)^2}{4\sigma_y^2}\right],
\label{eq:gauss_spec}
\end{equation}
where $q_x=2\pi/\ell_x$ and $q_y=2\pi/\ell_y$. For the JONSWAP spectrum we have
\begin{equation}
\psi(\vc k)
=\frac{\alpha^{1/2}}{k^2}\exp\left[-\frac{5}{8}\left(\frac{k_0}{k}\right)^2\right]\gamma^{r/2}\sqrt{D(\theta)},\quad
r = \exp \left[ -\frac{(\sqrt k - \sqrt{k_0})^2}{2\sigma^2k_0}\right].
\label{eq:jonswap}
\end{equation}
The directional spreading $D(\theta)$ is given by 
\begin{equation}
D(\theta)=
\begin{cases}
\frac{2}{\theta_0}\cos^2\left(\frac{\pi \theta}{\theta_0}\right), & \quad |\theta|\leq \frac{\theta_0}{2}\\
0, & \quad |\theta|> \frac{\theta_0}{2}
\end{cases},
\end{equation}
where $\theta$ is the propagation direction and the parameter $\theta_0$ is the directional spreading angle of the 
wave. 
\begin{table}[h]
\centering
\caption{List of parameters for the five sets of simulations considered here. 
For the Gaussian spectra we have $\sigma_x=0.035$ and  $\sigma_y=0.015$.
For the JONSWAP spectra we set $\gamma=3.3$, $\alpha=0.04$. The remaining 
parameters are listed in the table.
}
\begin{tabular}{c |c c| c c c}
 &  G1 & G2 & J1 & J2 & J3\\ \hline
Spectrum & Gaussian & Gaussian & Jonswap & Jonswap & Jonswap \\
Parameter & $\epsilon=0.035$ & $\epsilon=0.053$ & $\theta_0=40\degree$ & $\theta_0=60\degree$ & $\theta_0=80\degree$\\
\end{tabular}
\label{tab:params}
\end{table}

We consider five sets of experiments, two with the Gaussian spectrum and three with the
JONSWAP spectrum as listed in Table~\ref{tab:params}. For the Gaussian spectrum we set 
$\sigma_x=0.035$, $\sigma_y=0.015$ and consider 
two sets of values for the remaining parameter, $\epsilon=0.035$ (G1)
and $\epsilon=0.053$ (G2).
The standard deviations $\sigma_x$ and $\sigma_y$ are chosen such that
several wave groups fit in the computational domain. 
We choose $\sigma_y<\sigma_x$ since the waves tend to elongate orthogonal to the
propagation direction of the wave train (the $x$-axis here).
We also consider conservative values for the steepness $\epsilon$
to ensure the validity of the envelope equations~\eqref{eq:mnls_2d}.

For the JONSWAP spectrum, following~\cite{xiao2013}, we set $\sigma=0.07$ for $k\leq k_0$ 
and $\sigma=0.09$ for $k>k_0$. The peak enhancement factor is set to $\gamma=3.3$ 
to conform to experimentally measured spectra~\citep{hasselmann1973}. 
The amplitude is set to $\alpha=0.04$; this value is chosen so that
the resulting average wave height is around $0.05$, comparable to experiment G1.
We investigate the effect of the spreading angle since this is the parameter
that was absent in the
unidirectional context considered previously by~\cite{cousins16} and~\cite{PRE2016}.
As listed in Table~\ref{tab:params}, we consider two values of the spreading angle, 
$\theta_0=40\degree$ and $60\degree$  (experiments J1 and J2, respectively). For completeness  we also consider a third spectrum with spreading angle $\theta_0=80\degree$ (experiment J3). In this latter case, however, we have no rogue waves occurring due to modulation instability, but nevertheless we will include J3 in our results to demonstrate the trends as the spreading angle increases.   

\subsection{Group detection and predictions}\label{sec:PDFs}
For each experiment G1-G2 and J1-J3, 
we generate $10,000$ wave fields (note that the phases $\phi$ in~\eqref{eq:env_Fc} are random).
Using the group detection algorithm of Section~\ref{sec:decomp}, each wave field is decomposed into EWGs and
the length scales $L_x$, $L_y$ and amplitude $A_0$ of the wave groups are recorded. 

\begin{figure}
\centering
\subfloat[]{\includegraphics[width=.5\textwidth]{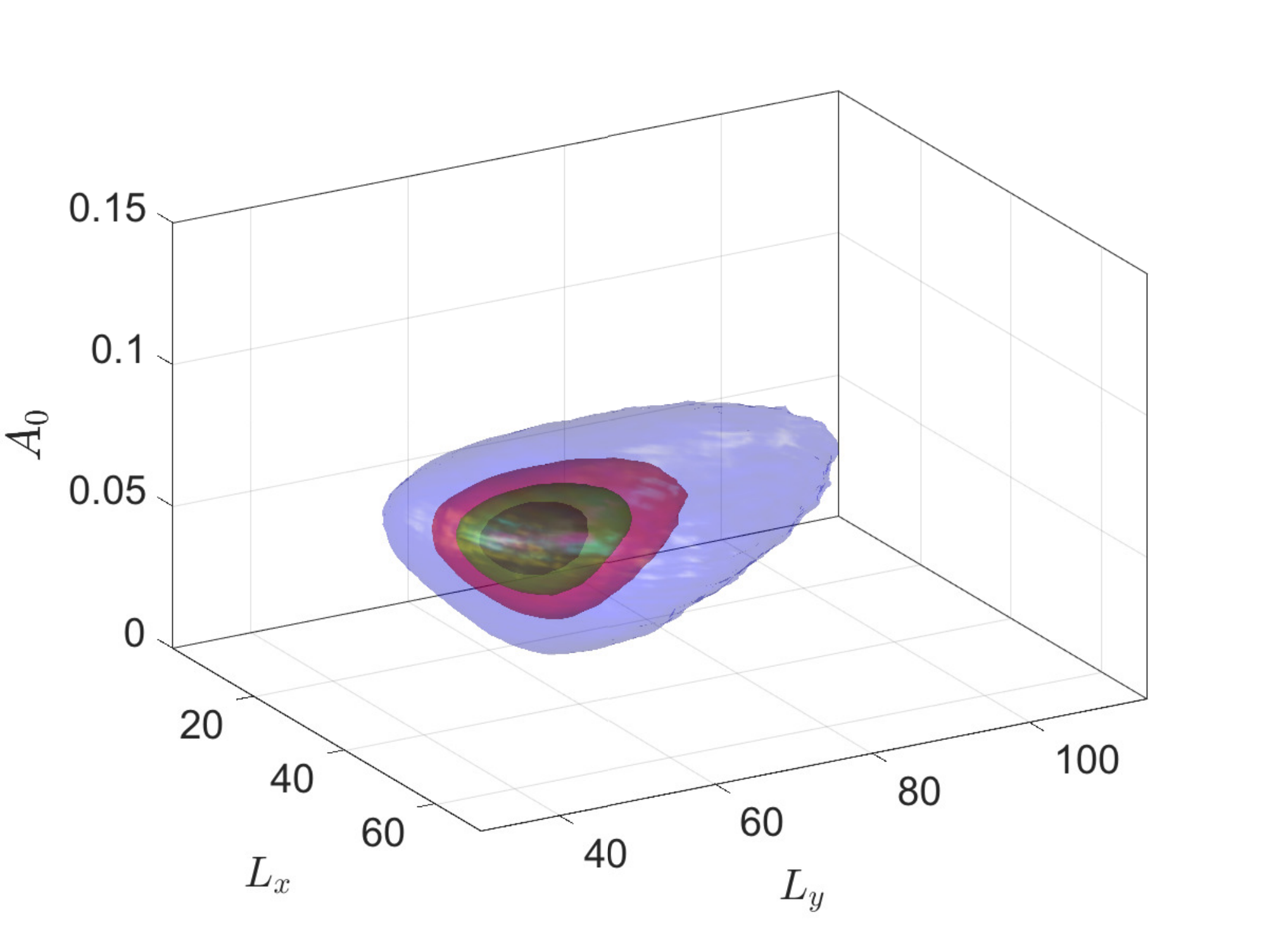}}
\subfloat[]{\includegraphics[width=.5\textwidth]{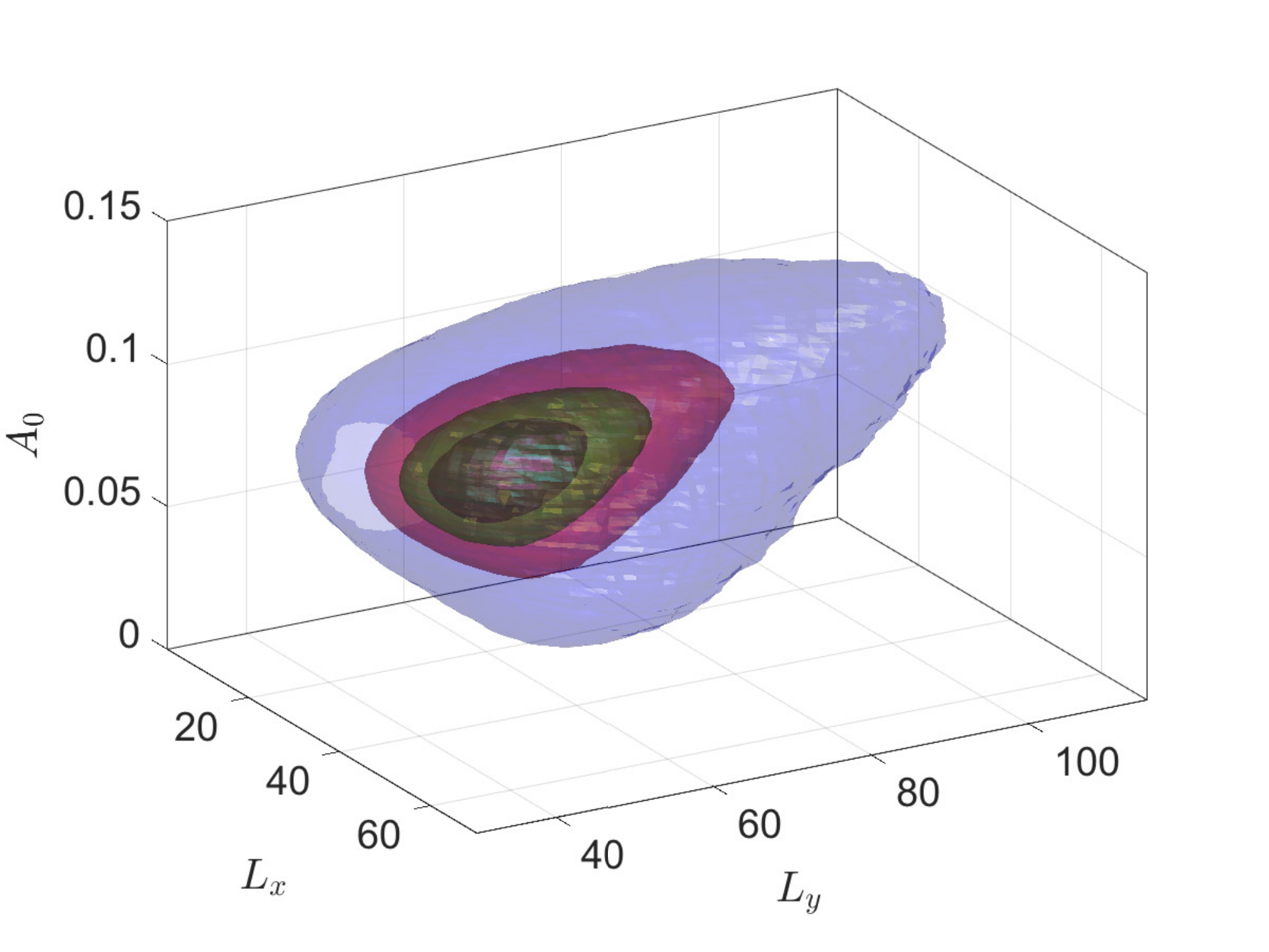}}
\caption{Joint PDF of the detected Gaussian wave groups with length scales $(L_x,L_y)$ and the
wave amplitude $A_0$. Each PDF is extracted from $10,000$ realizations of random waves with Gaussian
spectrum~\eqref{eq:gauss_spec}:
(a) G1, 
(b) G2. 
The figures shows $5$ isosurfaces of the PDFs, equispaced 
between $5\%$ and $95\%$ of the maximum PDF value.
}
\label{fig:groupDet_pdfIsoSurf}
\end{figure}

The resulting parameters $(L_x,L_y,A_0)$ span a wide range of values 
that depends on the shape of the spectrum.
Figure~\ref{fig:groupDet_pdfIsoSurf}, for instance, shows the joint probability distribution function (PDF) of the
wave group parameters $(L_x,L_y,A_0)$ resulting from the simulations G1 and G2.
As reported in Table~\ref{tab:stats}, the average peak height of the detected wave groups for experiments G1 and G2 are 
$\langle A_0\rangle =0.05$ and $\langle A_0\rangle =0.075$, respectively. The larger average peak height in G2 (compared to G1)
is expected as the parameter $\epsilon$ controls the average height of the resulting waves. 

We observe that the mean length scales $\langle L_x\rangle$ and $\langle L_y\rangle$ are very similar for the two
experiments G1 and G2 (see Table~\ref{tab:stats}). This is also expected as the standard deviations $\sigma_x$ and $\sigma_y$ are 
identical for the two spectra. The standard deviation of the length scales $L_x$ and $L_y$ are significantly
larger for the run G2 compared to G1. This is visibly appreciable from figure~\ref{fig:groupDet_pdfIsoSurf} that shows a broader
joint PDF for the run G2, spanning a larger range of length scales, which is a direct consequence of the larger energy content of the spectrum in this case.

Next we examine the predictive power of the EWG decomposition. To this end, we generate $1,000$
random wave field envelopes $u_0$ for each parameter set G1 and G2. Each envelope is evolved, using
the MNLS equation~\eqref{eq:mnls_2d}, for $T=1500$ time units and the maximum spatiotemporal amplitude of the 
field $\max_{x,y,t}|u(x,y,t)|$ is recorded. We refer to $\max_{x,y,t}|u(x,y,t)|$ as the \emph{true amplitude} of the wave.
\begin{table}[h!]
\centering
\caption{Statistics of the wave groups, rogue waves and their prediction. Here the wave period is $T_0=2\pi$
and the characteristic wave length is $\lambda_0=2\pi$.}
\begin{tabular}{c |c c| c c c}
 &  G1 & G2 & J1 & J2 &  J3\\ \hline
$\langle A_0\rangle$ & $0.050$ & $0.075$   & $0.054$ & $0.053$ & $0.052$ \\
$\langle L_x\rangle$ & $30.0$ & $30.2$     & $19.1$ & $18.9$ & $18.9$ \\
$\langle L_y\rangle$ & $67.7$ & $68.2$    & $53.8$ & $36.4$ & $28.3$ \\ 
$A_0^c$ & $0.087$ & $0.087$ & $0.092$ & $0.097$ & -- \\
$L_x^c$ & 32 & 31 & 20 & 21 & -- \\
$L_y^c$ & 76 & 76 & 63 & 47 & -- \\
\hline 
rogue waves  & $11.7\%$ & $25.6\%$ & $2.2\%$ & $1.4\%$ & $0\%$ \\
false negatives   &$4.9\%$ & $13.7\%$ & $1.4\%$ & $1.1\%$ & -- \\
false positives    & $3.9\%$ & $5.6\%$ & $0.2\%$ & $0.1\%$ & $0\%$\\
average warning time & $78T_0$ & $40T_0$  & $68T_0$ & $72T_0$ & --\\
average relative error  & $3.0\%$  & $9.2\%$ & $5.6\%$ & $5.0\%$ & $4.9\%$ 
\end{tabular}
\label{tab:stats}
\end{table}

\begin{figure}
\centering
\includegraphics[width=\textwidth]{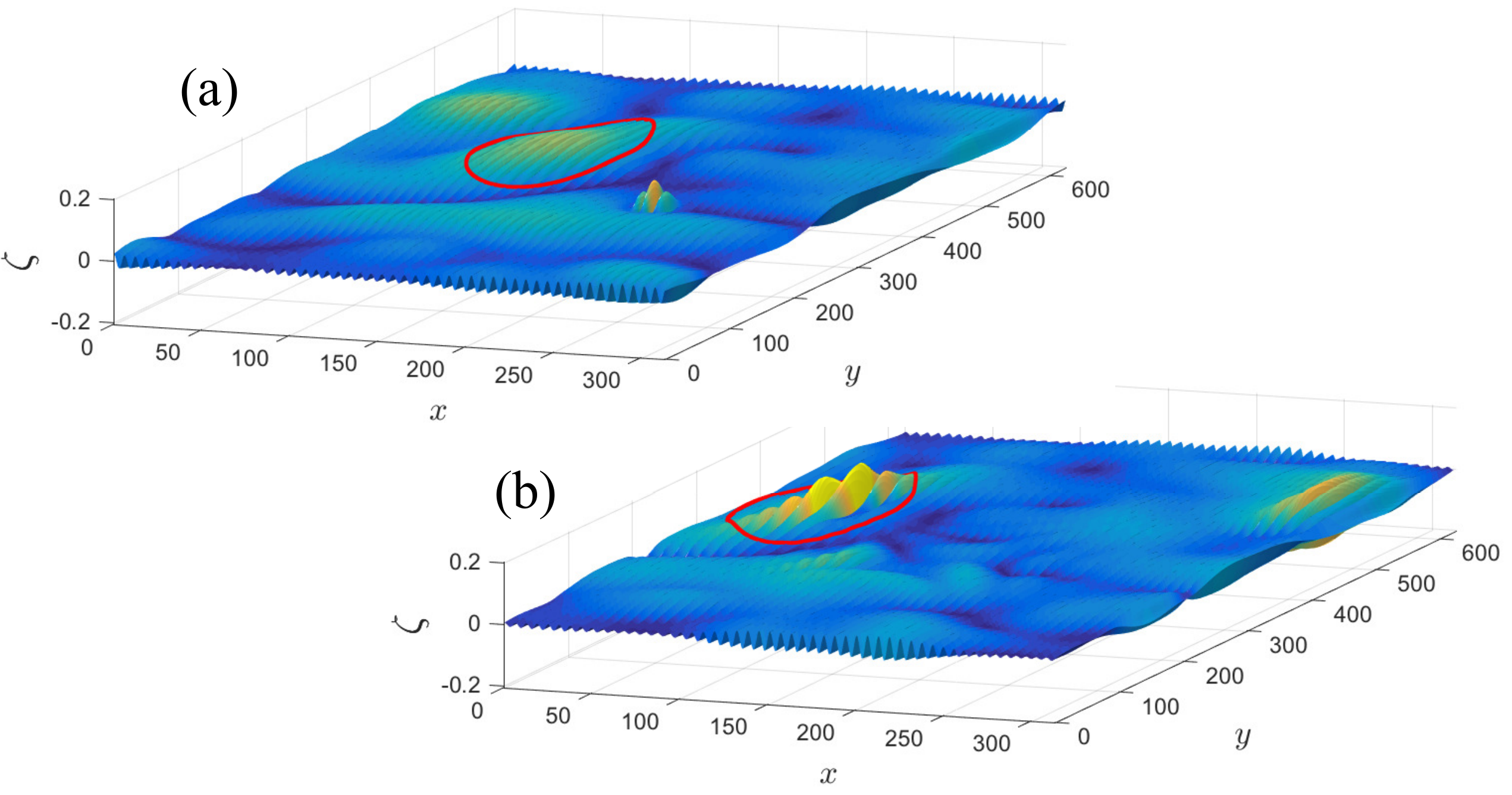}
\caption{A wave field generated from spectrum G2 at the initial time (a)
and 175 wave periods later (b).
The red curve marks the focusing wave group (identified at the initial time)
that develops into a rogue wave.}
\label{example_pred}
\end{figure}

\begin{figure}
\centering
\subfloat[]{\includegraphics[width=.45\textwidth]{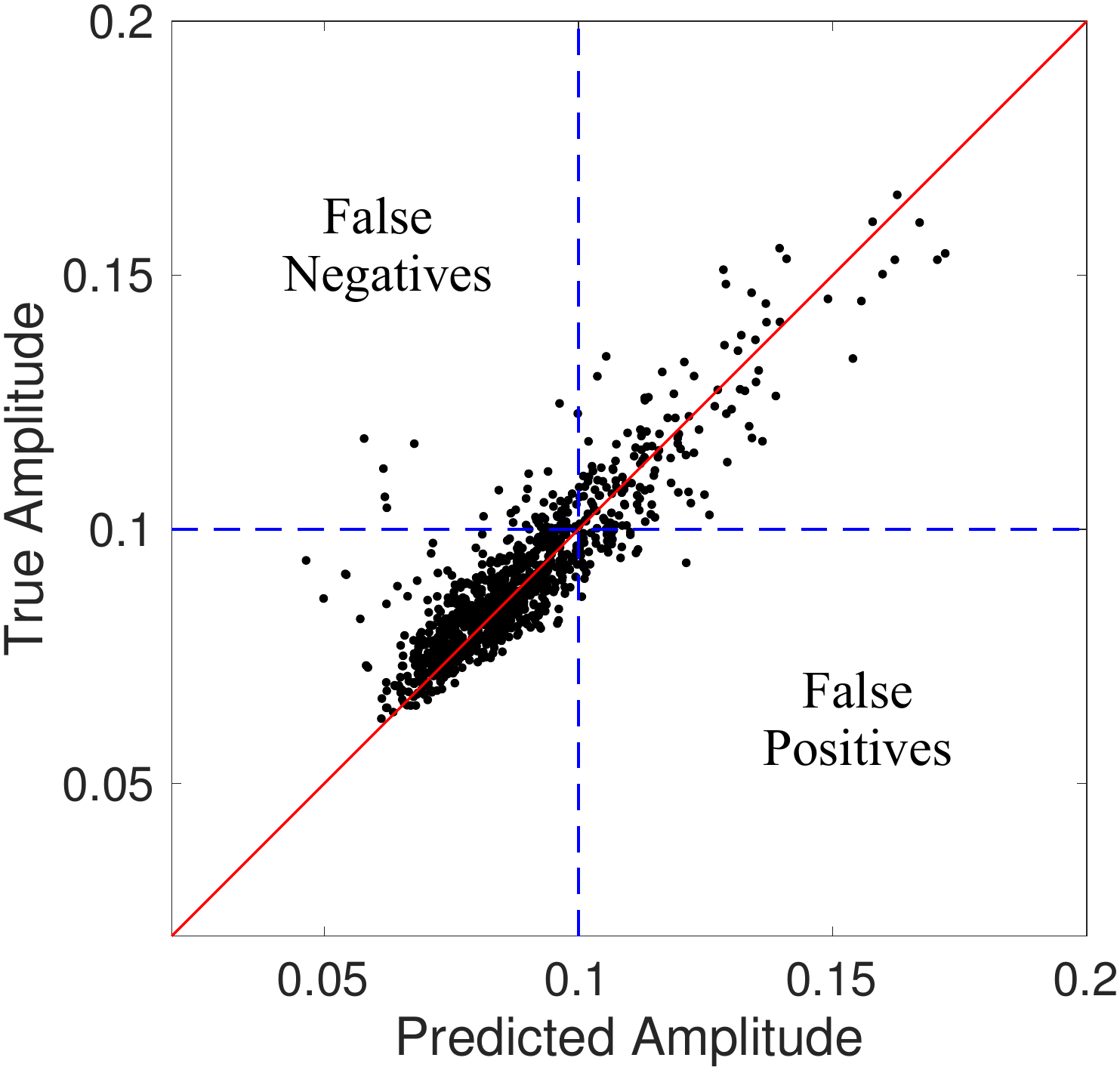}}
\subfloat[]{\includegraphics[width=.43\textwidth]{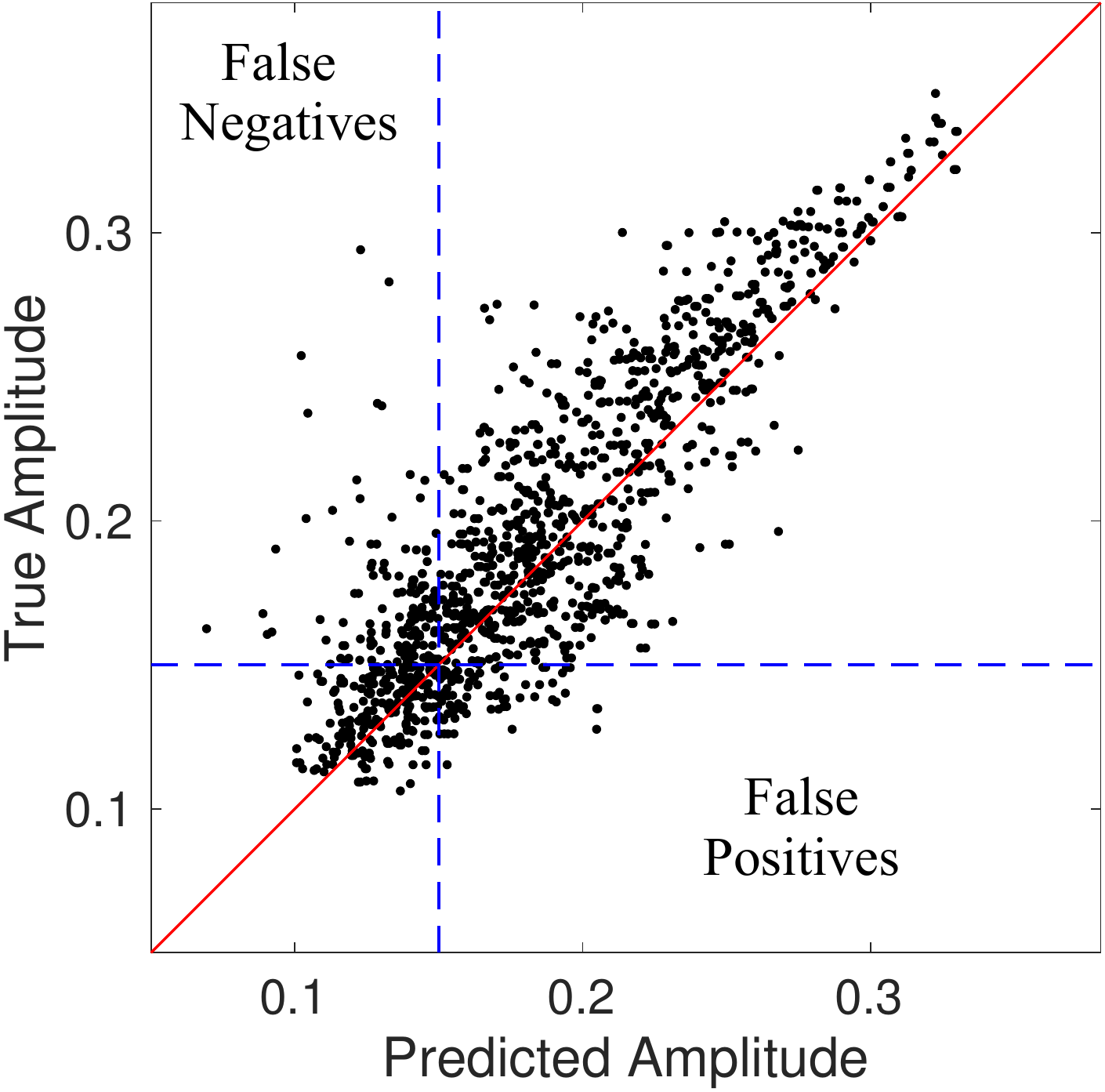}}
\caption{Predicted maximal amplitude $\max_{x,y,t}|u(x,y,t)|$ from Gaussian group detection
versus the true amplitude from direct numerical simulation of MNLS. 
(a) G1
(b) G2.
The blue dashed lines indicate the magnitude of a rogue wave.}
\label{fig:PredvsTruAmp}
\end{figure}

We predict the maximal amplitude of each wave field by decomposing it into EWGs 
and interpolating the function $\Amax$ that is precomputed from Section~\ref{sec:ewg}. 
We refer to the maximal resulting amplitude as the \emph{predicted amplitude} of the wave.
An example of such prediction is shown in Figure~\ref{example_pred}. The accuracy of the scheme is demonstrated in Figure~\ref{fig:PredvsTruAmp} that shows the true amplitudes versus the predicted amplitudes for the
experiments G1 and G2. The figures for the JONSWAP experiments J1-J3 are similar (not presented here).
As reported in Table~\ref{tab:stats}, on average, the relative errors of the predictions
are between $3\%$ and $9.2\%$. The relative prediction error is defined 
here as the ratio of the difference between the true amplitude and the predicted amplitude divided by the true amplitude:
\begin{equation}
\mbox{relative prediction error} = \frac{\big|\mbox{true amplitude}-\mbox{predicted amplitude}\big|}{\mbox{true amplitude}}.
\end{equation}
Given the approximations and assumptions underlying our reduced-order prediction, 
the resulting relative prediction errors of less than $9.2\%$ are quite satisfactory.


Now we consider the prediction of rogue waves using the reduced order method.
Following the convention, we defined an extreme (or rogue) wave as one whose 
height $H=2|u|$ exceeds twice the significant wave height $H_s$, where the
significant wave height is defined as four times the standard deviation of the 
surface elevation: $H_s = 4\sqrt{\langle \zeta^2\rangle}$.
The dashed blue lines in figure~\ref{fig:PredvsTruAmp} mark the resulting
rogue waves threshold, i.e., $|u|=H_s$. First, we observe that, compared to G1, a higher percentage of wave field from experiment G2
produce rogue waves (see Table~\ref{tab:stats}). This is to be expected as the 
spectral amplitude $\epsilon$ is larger in G2. As a result, the amplitude of individual
wave groups tend to be larger for G2. Another contributing factor
to this higher probability is the fact that the wave groups in G2 are more 
likely to have larger transverse length scales $L_y$ (cf. figure~\ref{fig:groupDet_pdfIsoSurf}).

\begin{figure}
\centering
\subfloat[]{\includegraphics[width=.33\textwidth]{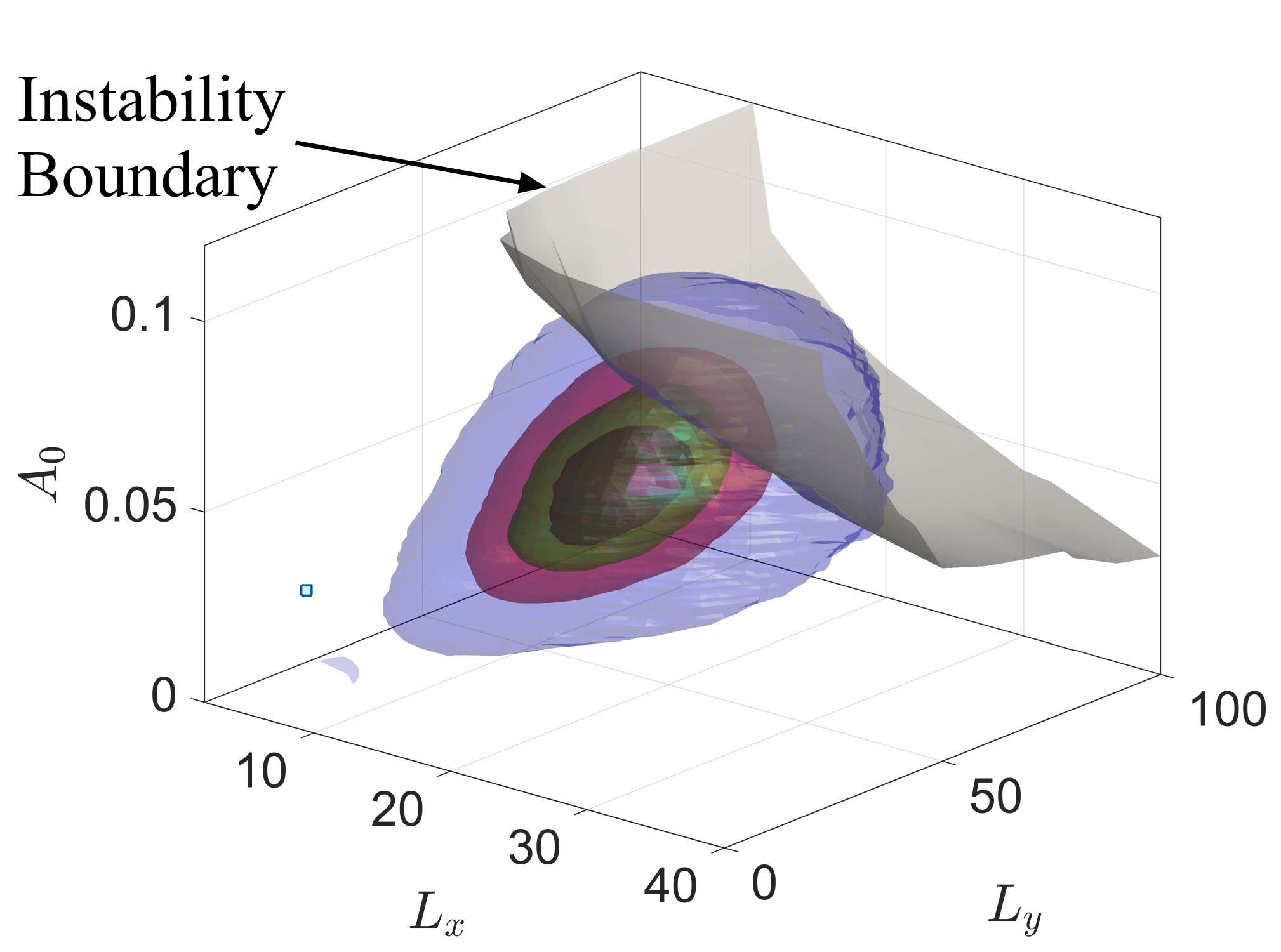}}
\subfloat[]{\includegraphics[width=.33\textwidth]{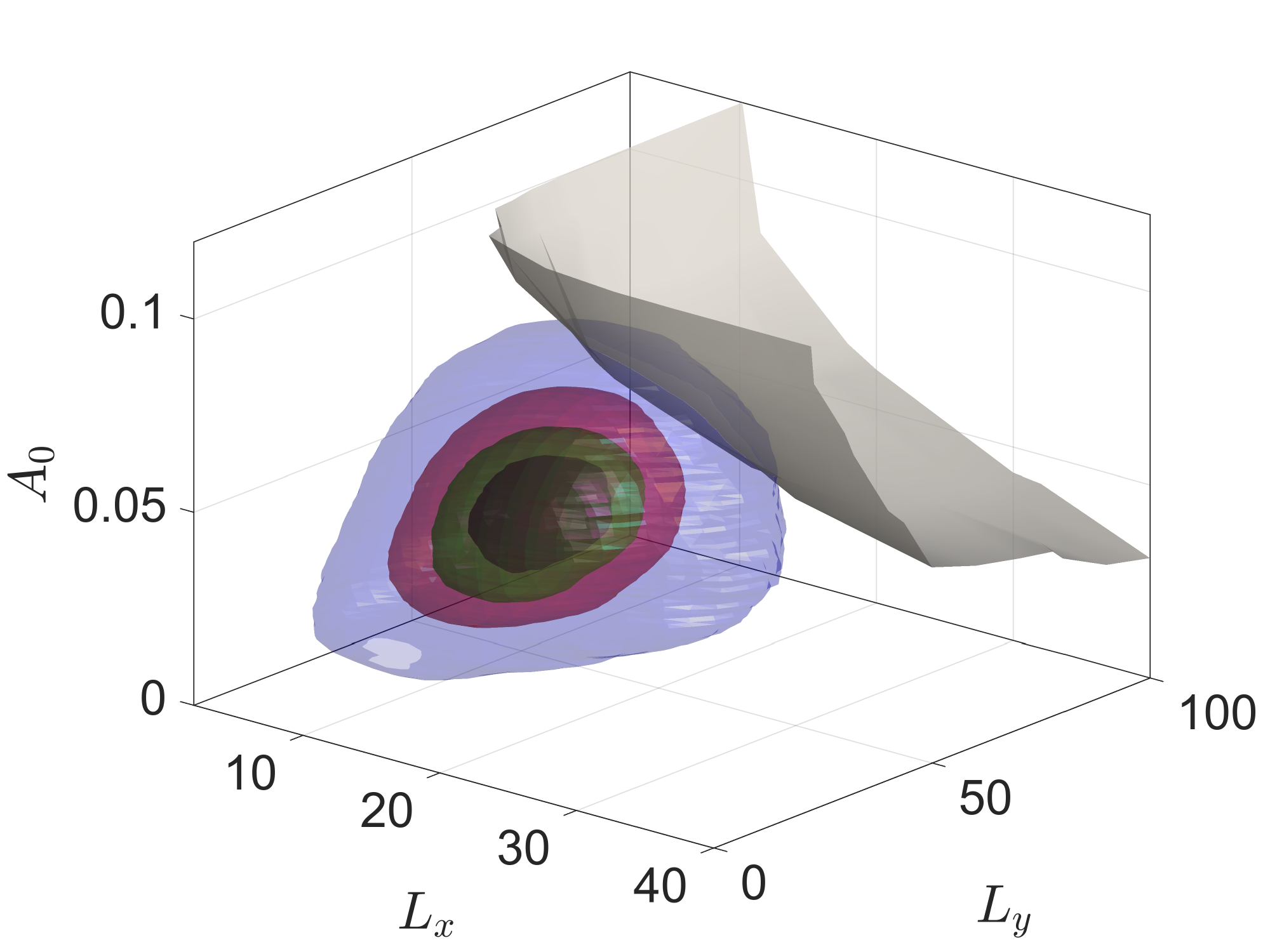}}
\subfloat[]{\includegraphics[width=.33\textwidth]{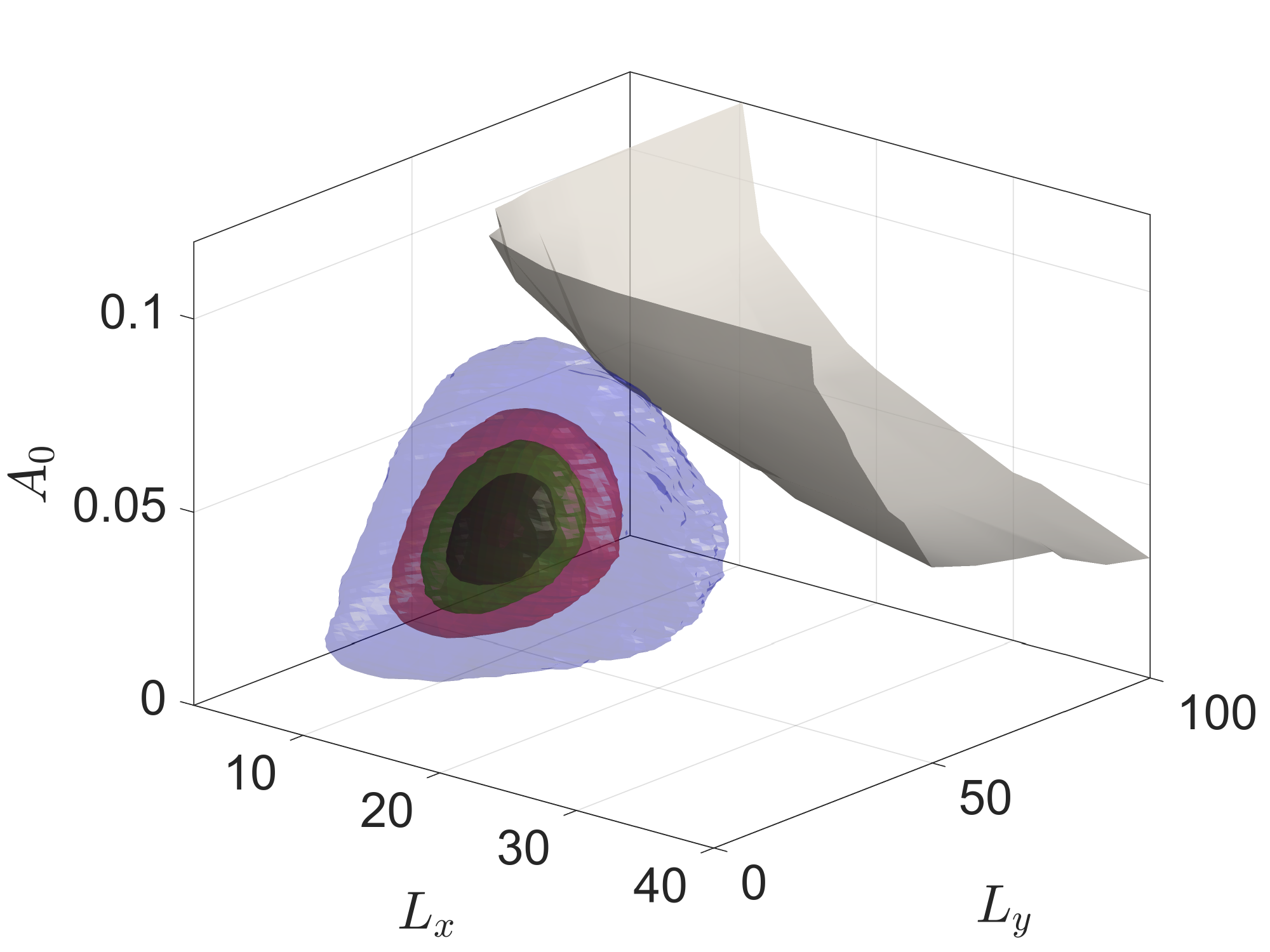}}
\caption{Joint PDF of the detected Gaussian wave groups with length scales $(L_x,L_y)$ and the
wave amplitude $A_0$. Each PDF is extracted from $10,000$ realizations of random waves with JONSWAP
spectrum with the spreading angle
$\theta_0=40\degree$ (a),  
$\theta_0=60\degree$ (b) and 
$\theta_0=80\degree$ (c). 
The figures shows $5$ isosurfaces of the PDFs, equispaced 
between $5\%$ and $95\%$ of the maximum PDF value.
The gray (dark) surface shows the instability boundary above which focusing instabilities form.
}
\label{fig:groupDet_pdfIsoSurf_jonswap}
\end{figure}

We now turn to the experiments J1-J3 with the JONSWAP spectrum~\eqref{eq:jonswap}. Here, we focus on the
directionality of the spectra controlled by the spreading angle $\theta_0$, keeping all the other parameters fixed. 
Figure~\ref{fig:groupDet_pdfIsoSurf_jonswap} shows the PDF of the
length scales of the detected EWGs obtained from $10,000$ randomly generated wave fields. 
The average wave amplitudes $\langle A_0\rangle$ and lengths scales $\langle L_x\rangle$ and $\langle L_y\rangle$
are reported in Table~\ref{tab:stats}. We first observe that the average wave amplitudes $\langle A_0\rangle\approx 0.05$ 
are similar for the three experiments J1-J3. In spite of this similarity, the frequency of
rogue wave occurrence decreases as the spreading angle $\theta_0$ increases. For the 
largest angle $\theta_0=80$, for instance, no rogue waves were produced from the $1000$
wave fields considered. This is a well-known effect reported previously in several numerical and experimental studies (see, e.g., 
\cite{Onorato02,xiao2013}).
\begin{figure}
\centering
\subfloat[]{\includegraphics[width=.5\textwidth]{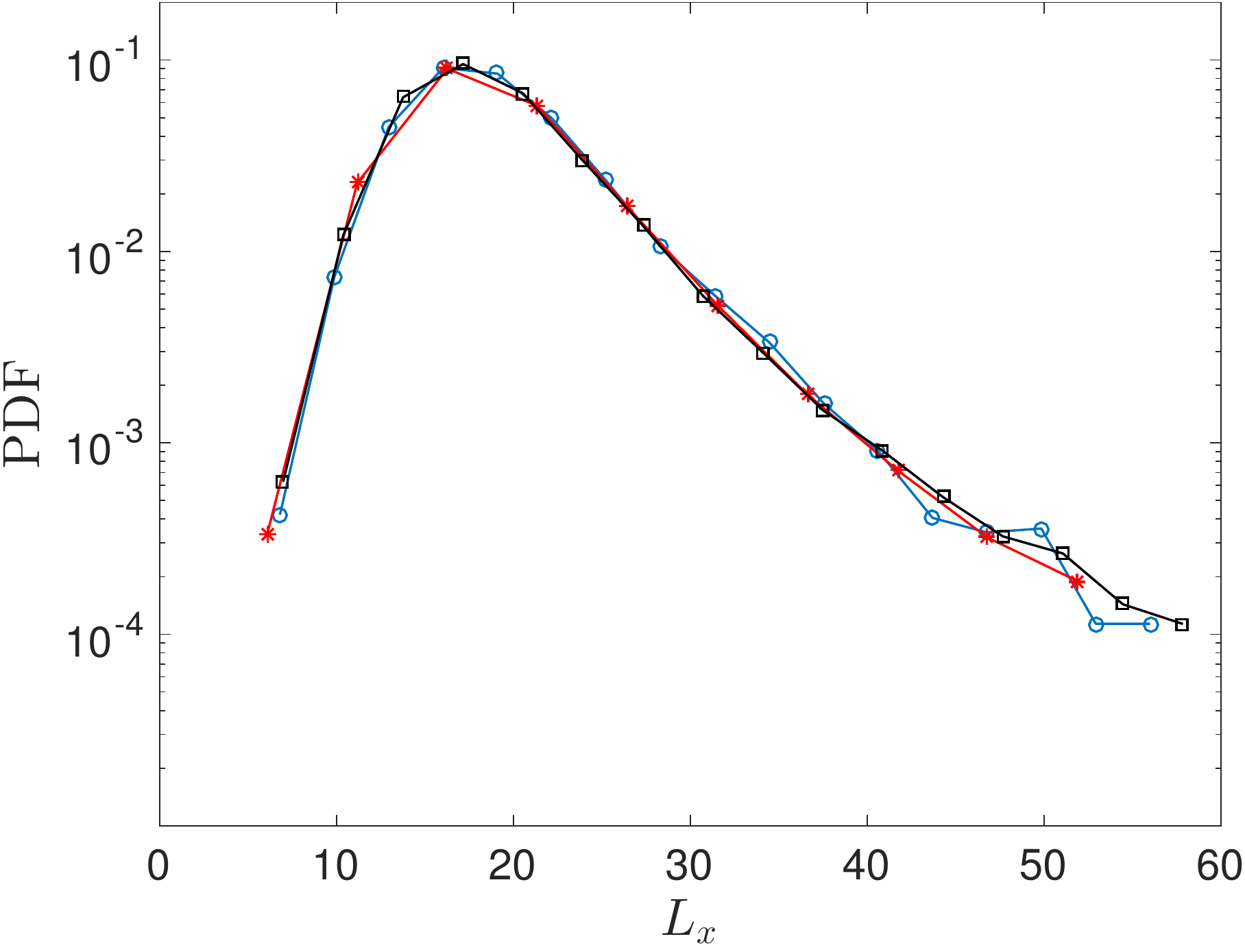}}
\subfloat[]{\includegraphics[width=.5\textwidth]{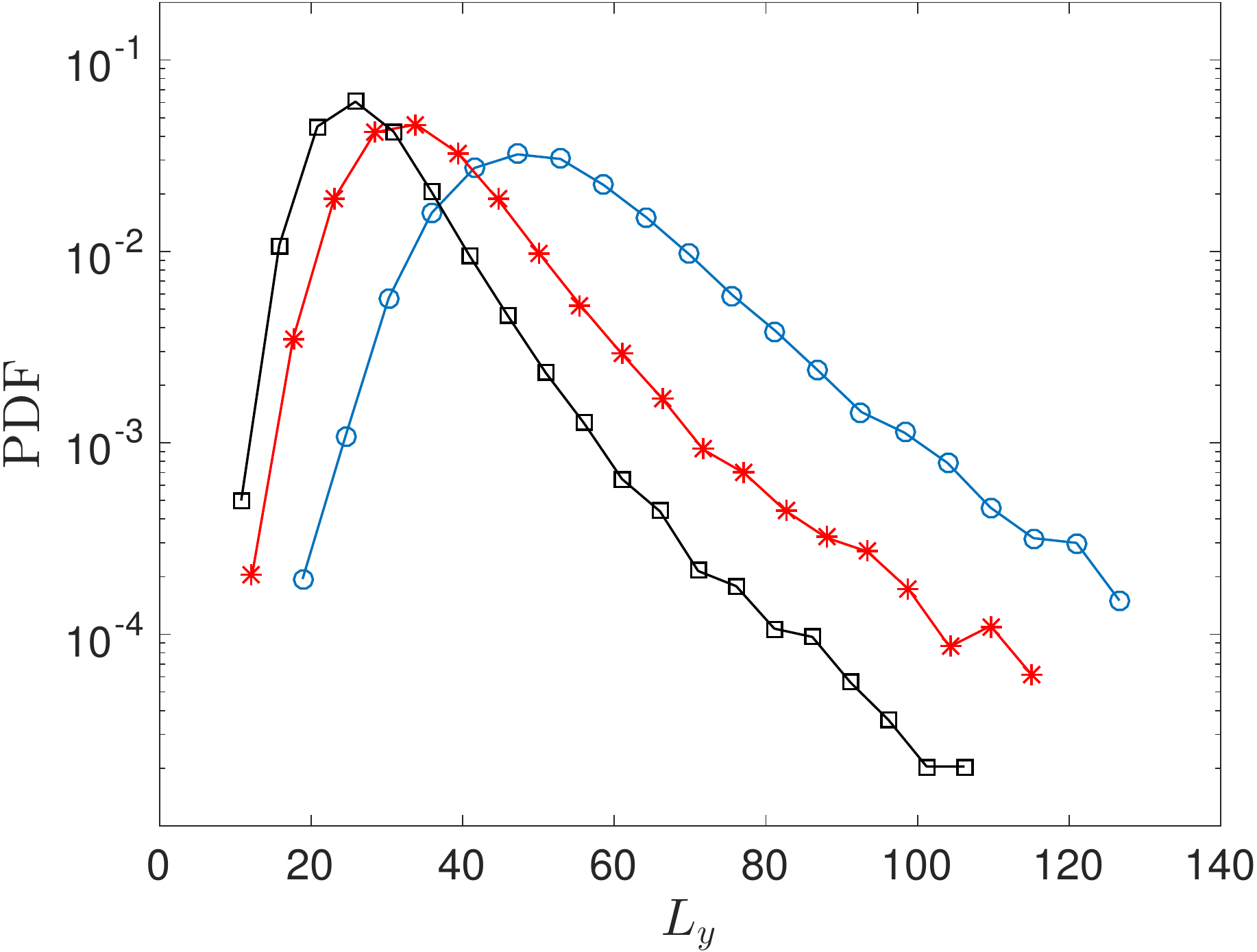}}
\caption{PDF of the detected length scales $L_x$ and $L_y$ from JONSWAP simulations
J1 (blue, circles),
J2 (red, stars) and
J3 (black, square).
}
\label{fig:pdfL_jonswap}
\end{figure}

The reason for this decreasing probability is clear from our wave group analysis presented in 
figure~\ref{fig:groupDet_pdfIsoSurf_jonswap}. As the spreading angle $\theta_0$ increases the entire
PDF shifts towards lower values of $L_y$, i.e. the 
average transversal length scale $\langle L_y\rangle$ decreases. 
This is better seen in the one-dimensional margnial PDFs shown in figure~\ref{fig:pdfL_jonswap}.
Recall from Section~\ref{sec:ewg}
that the wave groups with smaller $L_y$ require larger initial amplitude $A_0$ in order to focus
(cf. figure~\ref{fig:instBoundary}). The wave fields with larger spreading angle tend to have 
wave groups with smaller transversal length scales $L_y$, while having similar wave amplitudes $A_0$. 
As a result, they are less likely to produce rogue waves. We point out that the
distribution of the longitudinal length scales $L_x$ are quite insensitive to the
spreading angle $\theta_0$ (see figure~\ref{fig:pdfL_jonswap}(a)).

We point out that the computational time required by the reduced-order prediction is
significantly shorter than the time required for evolving the wave field under the
envelope equation. For instance, evolving the wave fields under the MNLS equation
for $100$ wave periods (using the ETD4RK scheme with the time-step size $\Delta t=0.025$ and $2^9\times 2^8$ grid points)
takes approximately $1800$ seconds ($= 30$ minutes). Decomposing the same wave fields
into the Gaussian EWGs, on the other hand, takes between $10$ and $40$ seconds. 
The variation in the decomposition time is due to the variations in the number of 
wave groups $N$ (cf. Eq.~\eqref{eq:gauss_series}) that are
present in different randomly generated wave fields.

\subsection{Critical length scales and amplitudes}
Recall from Section~\ref{sec:ewg} that particular combinations of the length scales 
$(L_x,L_y)$ and the amplitude $A_0$ are required for the wave group to 
evolve into a rogue wave. Then the natural question is: Given a particular wave spectrum, what 
critical combination $(L_x,L_y,A_0)$ is most likely to produce rogue waves?
The EWG evolution (figure~\ref{fig:instBoundary}) together with the PDFs of the detected groups
(figures~\ref{fig:groupDet_pdfIsoSurf} and~\ref{fig:groupDet_pdfIsoSurf_jonswap}) 
are sufficient to answer this question.

For a given wave spectrum, let $\Aext$ denote the amplitude threshold for a rogue wave. 
That is a wave with amplitude $|u|>\Aext$ constitutes a rogue wave. 
Following the conventional definition of a rogue wave (waves with height greater than twice the
significant wave height $H_s$), we have $\Aext=H_s$.
Considering the maximum amplitude function $\Amax$ shown in figure~\ref{fig:instBoundary},
the rogue waves lie above the critical surface
\begin{equation}
\Amax(L_x,L_y,A_0) =  \Aext.
\end{equation}
Since this critical surface is a graph over $(L_x,L_y)$ variables, the implicit function theorem~\citep{rudin64} 
guarantees that there is a function $A_0^c:\mathbb R\times \mathbb R\to\mathbb R$ such that
\begin{equation}
\Amax\big(L_x,L_y,A_0^c(L_x,L_y)\big) =  \Aext,
\label{eq:critLength}
\end{equation}
for all $L_x$ and $L_y$. The quantity $A_0^c(L_x,L_y)$ is the minimal initial amplitude required for 
an EWG with length scales $(L_x,L_y)$ to develop into a rogue wave at some point in the future.

\begin{figure}
\centering
\includegraphics[width=.9\textwidth]{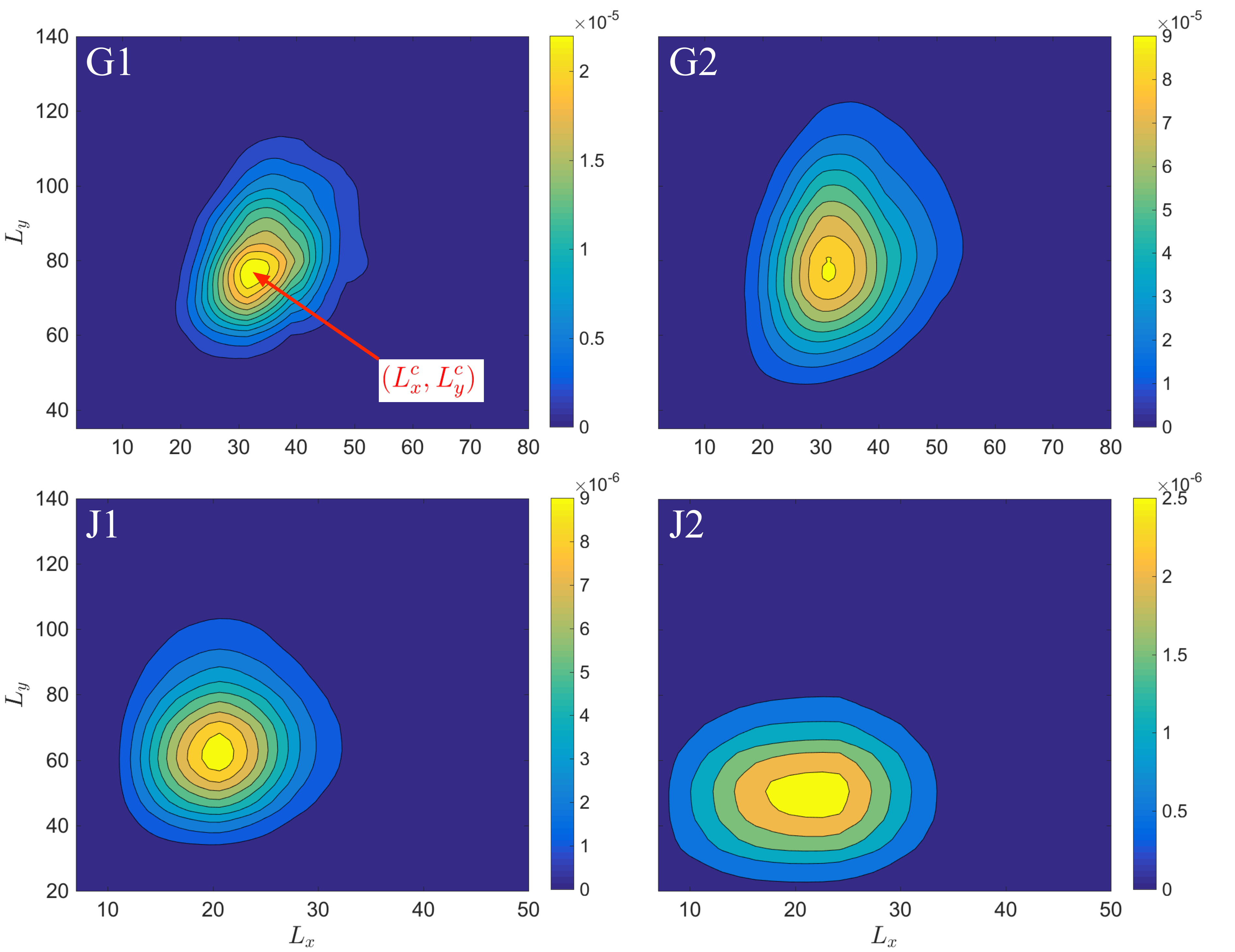}
\caption{The probability of rogue wave formation $\Pext$~\eqref{eq:Pext} as a function of the length scales $L_x$
and $L_y$ for the experiments G1-G2 and J1-J2. The probability for J3 is negligible and hence is not shown.
The peak of $\Pext$ marks the critical length scales $(L_x^c,L_y^c)$.
}
\label{fig:Pext}
\end{figure}

On the other hand, for a given wave spectrum, the probability distribution function $p(L_x,L_y,A_0)$
is computed in Section~\ref{sec:PDFs} (figures~\ref{fig:groupDet_pdfIsoSurf} and~\ref{fig:groupDet_pdfIsoSurf_jonswap}).
Using this three-dimensional PDF, we define the conditional probability of a rogue wave with the given length scales $L_x$ and 
$L_y$ as 
\begin{equation}
\Pext(L_x,L_y)=\int_{A_0^c(L_x,L_y)}^{\infty}p(L_x,L_y,A_0)\id A_0,
\label{eq:Pext}
\end{equation}
where $A_0^c$ is the critical amplitude defined in equation~\eqref{eq:critLength}.

Figure~\ref{fig:Pext} shows the probability $\Pext$ for the energy spectra listed in Table~\ref{tab:params}.
Each probability function has a distinct peak at some length scales $(L_x^c,L_y^c)$ reported in Table~\ref{tab:stats}.
We denote the associated critical amplitude $A_0^c(L_x^c,L_y^c)$ by $A_0^c$ for simplicity.
The physical interpretation of the triplet $(L_x^c,L_y^c,A_0^c)$ is the following. For a given spectrum, 
rogue waves most likely develop from wave groups whose length scales are $(L_x^c,L_y^c)$ and their 
amplitudes are larger than or equal to $A_0^c$. Roughly speaking, this means that 
the wave groups with length scales $(L_x^c,L_y^c)$ and amplitudes larger than $A_0^c$ are 
the `most dangerous' both in terms of dynamics (i.e. they tend to evolve to rogue waves) and also in terms of probability of occurrence.
 
The critical values  $(L_x^c,L_y^c,A_0^c)$ are listed in Table~\ref{tab:stats} for each wave spectrum. 
Focusing on the spectra G1 and G2, we observe that the critical length scale $L^c_x\simeq 31$ is
quite close to the average longitudinal lengths $\langle L_x\rangle\simeq 30$.
In contrast, the critical length $L_y^c\simeq 76$ is significantly larger than the 
typical transverse length scale $\langle L_y\rangle \simeq 68$. The critical amplitude
$A_0^c\simeq 0.087$ is also similar for the two spectra. The crucial difference is however 
the fact that the average amplitude of the wave groups
for G2 ($\langle A_0\rangle=0.075$) is significantly larger than G1 ($\langle A_0\rangle=0.05$). As a result, 
the wave groups from G2 are more likely to achieve the critical amplitude for rogue wave formation. 
This explains the high rate of rogue wave formation ($25.6\%$) observed for G2.
Similar observations can be made about the spectra J1 and J2.

\section{Summary and Conclusion}

Large waves form as a result of the dispersive mixing of smaller wave groups and the subsequent focusing due to nonlinear effects. Here we 
provide a method to quantify this mechanism in a reduced-order fashion. 
Specifically, we first develop a wave field decomposition algorithm that robustly and efficiently
represents a wave field as a discrete set of elementary wave groups
(EWGs) with Gaussian profiles. We then utilize the governing envelope equations to quantify the evolution of each of those elementary wave groups.

The combination of the wave field decomposition
algorithm, which provides information about the statistics of the system (caused by the dispersive mixing of harmonics),
and the precomputed map for the expected wave group elevation, which  encodes
dynamical information for focusing phenomena, allows for (i) the  understanding of how the probability
of occurrence of rogue waves changes as the spectrum parameters vary, (ii)
the computation of a critical lengthscale characterizing wave groups with
high probability of evolving to rogue waves, and (iii) the formulation of
a robust and parsimonious reduced-order prediction scheme for large waves. 

Specifically, the EWG decomposition combined with the precomputed map provides information that complements the usual spectral
analysis of the wave 
field, when it comes to large waves. For instance, it has been reported that
the frequency of rogue wave occurrence is decreasing as the spreading angle increases. Our EWG analysis reveals that
larger spreading angles lead to wave groups whose width in the transverse
direction,  $L_y$, 
tends to be smaller. This has a direct interpretation in terms of the probability of occurrence of rogue waves.

Through the same analysis we also identified a critical lengths scale for each direction  $(L_x^c,L_y^c)$ and a critical amplitude
$A_0^c$ associated with rogue waves. 
For each wave spectrum these are the most likely combination of scales and
amplitudes
that will eventually grow into a rogue wave. We found that the conditional
probability of 
the critical length scales $(L_x^c,L_y^c)$ is directly related to the frequency
of rogue wave occurrence.

Regarding the prediction scheme, we showed, through extensive direct numerical simulations of the
modified nonlinear Schr\" odinger (MNLS) equation, that the proposed reduced-order
method is capable of predicting the future wave height
with less than $10\%$ relative error and a rogue-wave-prediction time window between
$40$ and $78$ wave periods. The scheme is orders of magnitude less expensive compared with direct simulation methods and it is very robust with respect to measurement noise, which is inevitable in any realistic setting.

Our work can be extended in several directions. An important direction is the case of crossing seas where wave groups can carry more than one dominant wavenumber. Other scenarios where our analysis can be extended in a straightforward manner is the case of wave-current interaction and finite bathymetry. For such cases we expect that the stability and response surfaces will be modified in order to take into account the additional effect.
In such cases, envelope equations may not be the best approach to characterize the dynamics of EWGs and more direct methods should be utilized. Clearly an important step forward is the experimental validation of the proposed scheme and we currently work towards this direction. The proposed approach provides an important paradigm of how the combination of dynamics and statistics can lead to better understanding of the system properties. In addition, it paves the way for the design of practical and robust prediction systems for large waves in the ocean.

\section*{Acknowledgments}
T.P.S. has been supported through the ONR\ grants N00014-14-1-0520 and N00014-15-1-2381 and the AFOSR grant FA9550-16-1-0231. M.F.
has been supported through the second grant. We are also grateful to the American Bureau of Shipping for support under a Career Development Chair at MIT.\appendix


\begin{thebibliography}{}

\bibitem[Adcock et~al., 2012]{Adcock12}
Adcock, T. A.~A., Gibbs, R.~H., and Taylor, P.~H. (2012).
\newblock The nonlinear evolution and approximate scaling of directionally
  spread wave groups on deep water.
\newblock {\em Proc. R. Soci. A}, 468(2145):2704--2721.

\bibitem[Benjamin and Feir, 1967]{benjamin67}
Benjamin, T.~B. and Feir, J.~E. (1967).
\newblock The disintegration of wave trains on deep water part 1. theory.
\newblock {\em J. Fluid Mech.}, 27(03):417--430.

\bibitem[Bishop, 1995]{bishop1995}
Bishop, C.~M. (1995).
\newblock {\em Neural networks for pattern recognition}.
\newblock Oxford university press.

\bibitem[Borge et~al., 2013]{borge2013}
Borge, J. C.~N., Reichert, K., and Hessner, K. (2013).
\newblock Detection of spatio-temporal wave grouping properties by using
  temporal sequences of {X}-band radar images of the sea surface.
\newblock {\em Ocean Modelling}, 61:21--37.

\bibitem[Chabchoub, 2016]{chabchoub2016}
Chabchoub, A. (2016).
\newblock Tracking breather dynamics in irregular sea state conditions.
\newblock {\em Phys. Rev. Lett.}, 117:144103.

\bibitem[Chabchoub et~al., 2011]{chabchoub2011}
Chabchoub, A., Hoffmann, N.~P., and Akhmediev, N. (2011).
\newblock Rogue wave observation in a water wave tank.
\newblock {\em Phys. Rev. Lett.}, 106(20):204502.

\bibitem[Clauss et~al., 2014]{Clauss2014}
Clauss, G.~F., Klein, M., Dudek, M., and Onorato, M. (2014).
\newblock {Application of Higher Order Spectral Method for Deterministic Wave
  Forecast}.
\newblock In {\em Volume 8B: Ocean Engineering}, page V08BT06A038. ASME.

\bibitem[Cousins and Sapsis, 2014]{cousins14}
Cousins, W. and Sapsis, T.~P. (2014).
\newblock Quantification and prediction of extreme events in a one-dimensional
  nonlinear dispersive wave model.
\newblock {\em Physica D}, 280:48--58.

\bibitem[Cousins and Sapsis, 2015]{cousins15}
Cousins, W. and Sapsis, T.~P. (2015).
\newblock Unsteady evolution of localized unidirectional deep-water wave
  groups.
\newblock {\em Phys. Rev. E}, 91(6):063204.

\bibitem[Cousins and Sapsis, 2016]{cousins16}
Cousins, W. and Sapsis, T.~P. (2016).
\newblock Reduced-order precursors of rare events in unidirectional nonlinear
  water waves.
\newblock {\em J. Fluid Mech}, 790:368--388.

\bibitem[Cox and Matthews, 2002]{cox02}
Cox, S. and Matthews, P. (2002).
\newblock Exponential time differencing for stiff systems.
\newblock {\em Journal of Computational Physics}, 176(2):430--455.

\bibitem[Craig et~al., 2012]{Craig2012}
Craig, W., Guyenne, P., and Sulem, C. (2012).
\newblock Hamiltonian higher-order nonlinear schrödinger equations for
  broader-banded waves on deep water.
\newblock {\em European Journal of Mechanics - B/Fluids}, 32:22 -- 31.

\bibitem[Dommermuth and Yue, 1987]{dommermuth1987}
Dommermuth, D.~G. and Yue, D. K.~P. (1987).
\newblock {A high-order spectral method for the study of nonlinear gravity
  waves}.
\newblock {\em Journal of Fluid Mechanics}, 184:267--288.

\bibitem[Dysthe et~al., 2008]{dysthe08}
Dysthe, K., Krogstad, H.~E., and M{\"u}ller, P. (2008).
\newblock Oceanic rogue waves.
\newblock {\em Annu. Rev. Fluid Mech.}, 40:287--310.

\bibitem[Dysthe, 1979]{dysthe79}
Dysthe, K.~B. (1979).
\newblock Note on a modification to the nonlinear {S}chr{\" o}dinger equation
  for application to deep water waves.
\newblock {\em Proc. R. Soc. A}, 369(1736):105--114.

\bibitem[Dysthe et~al., 2003]{dysthe03}
Dysthe, K.~B., Trulsen, K., Krogstad, H.~E., and Socquet-Juglard, H. (2003).
\newblock Evolution of a narrow-band spectrum of random surface gravity waves.
\newblock {\em J. Fluid Mech.}, 478:1--10.

\bibitem[Farazmand, 2016]{faraz_adjoint}
Farazmand, M. (2016).
\newblock An adjoint-based approach for finding invariant solutions of
  {N}avier-{S}tokes equations.
\newblock {\em J. Fluid Mech.}, 795:278--312.

\bibitem[Farazmand and Sapsis, 2016]{PRE2016}
Farazmand, M. and Sapsis, T.~P. (2016).
\newblock Dynamical indicators for the prediction of bursting phenomena in
  high-dimensional systems.
\newblock {\em Phys. Rev. E}, 94:032212.

\bibitem[Fedele et~al., 2011]{fedele2011}
Fedele, F., Benetazzo, A., and Forristall, G.~Z. (2011).
\newblock Space-time waves and spectra in the northern adriatic sea via a wave
  acquisition stereo system.
\newblock In {\em ASME 2011 30th International Conference on Ocean, Offshore
  and Arctic Engineering}, pages 651--663.

\bibitem[Forristall, 2000]{forristall2000}
Forristall, G.~Z. (2000).
\newblock Wave crest distributions: {O}bservations and second-order theory.
\newblock {\em J. Phys. Oceanogr.}, 30(8):1931--1943.

\bibitem[Friedman et~al., 2001]{friedman2001}
Friedman, J., Hastie, T., and Tibshirani, R. (2001).
\newblock {\em The elements of statistical learning}, volume~1.
\newblock Springer series in statistics Springer, Berlin.

\bibitem[Fu et~al., 2011]{fu_2011}
Fu, T.~C., Fullerton, A.~M., Hackett, E.~E., and Merrill, C. (2011).
\newblock {Shipboard measurments of ocean waves}.
\newblock In {\em OMAE 2011}, pages 1--8.

\bibitem[Gramstad and Trulsen, 2011]{gramstad2011}
Gramstad, O. and Trulsen, K. (2011).
\newblock Hamiltonian form of the modified nonlinear schr{\" o}dinger equation
  for gravity waves on arbitrary depth.
\newblock {\em J. Fluid Mech.}, 670:404--426.

\bibitem[Hasselmann et~al., 1973]{hasselmann1973}
Hasselmann, K., Barnett, T., Bouws, E., Carlson, H., Cartwright, D., Enke, K.,
  Ewing, J., Gienapp, H., Hasselmann, D., Kruseman, P., et~al. (1973).
\newblock Measurements of wind-wave growth and swell decay during the joint
  north sea wave project (jonswap).
\newblock Technical report, Deutches Hydrographisches Institut.

\bibitem[Janssen, 2003]{janssen03}
Janssen, P. A. E.~M. (2003).
\newblock Nonlinear four-wave interactions and freak waves.
\newblock {\em Journal of Physical Oceanography}, 33(4):863--884.

\bibitem[Longuet-Higgins, 1952]{longuet1952}
Longuet-Higgins, M.~S. (1952).
\newblock On the statistical distribution of the heights of sea waves.
\newblock {\em J. Mar. Res.}, 11(3):245--266.

\bibitem[Mei et~al., 2005]{mei2005}
Mei, C.~C., Stiassnie, M., and Yue, D. K.-P. (2005).
\newblock {\em Theory and applications of ocean surface waves: nonlinear
  aspects}, volume~23.

\bibitem[Nieto~Borge et~al., 2013]{borge13}
Nieto~Borge, J.~C., Reichert, K., and Hessner, K. (2013).
\newblock Detection of spatio-temporal wave grouping properties by using
  temporal sequences of {X}-band radar images of the sea surface.
\newblock {\em Ocean Modelling}, 61:21--37.

\bibitem[Nieto~Borge et~al., 2004]{nieto04}
Nieto~Borge, J.~C., Rodr{\'I}guez, G.~R., Hessner, K., and Gonz{\'a}lez, P.~I.
  (2004).
\newblock Inversion of marine radar images for surface wave analysis.
\newblock {\em Journal of Atmospheric and Oceanic Technology},
  21(8):1291--1300.

\bibitem[Onorato et~al., 2002]{Onorato02}
Onorato, M., Osborne, A.~R., and Serio, M. (2002).
\newblock Extreme wave events in directional, random oceanic sea states.
\newblock {\em Phys. Fluids}, 14(4):L25--L28.

\bibitem[Onorato et~al., 2004]{onorato2004}
Onorato, M., Osborne, A.~R., Serio, M., Cavaleri, L., Brandini, C., and
  Stansberg, C.~T. (2004).
\newblock Observation of strongly non-{G}aussian statistics for random sea
  surface gravity waves in wave flume experiments.
\newblock {\em Phys. Rev. E}, 70(6):067302.

\bibitem[Onorato et~al., 2013]{Onorato13}
Onorato, M., Residori, S., Bortolozzo, U., Montina, A., and Arecchi, F.~T.
  (2013).
\newblock Rogue waves and their generating mechanisms in different physical
  contexts.
\newblock {\em Physics Reports}, 528(2):47 -- 89.

\bibitem[Pierre, 1969]{pierre1969}
Pierre, D.~A. (1969).
\newblock {\em Optimization theory with applications}.
\newblock Dover Publishing, Inc.

\bibitem[Press et~al., 2007]{recipe}
Press, W.~H., Teukolsky, S.~A., Vetterling, W.~T., and Flannery, B.~P. (2007).
\newblock {\em Numerical recipes: {T}he art of scientific computing}.
\newblock Cambridge university press, third edition.

\bibitem[Ruban, 2015a]{ruban2015}
Ruban, V.~P. (2015a).
\newblock Anomalous wave as a result of the collision of two wave groups on the
  sea surface.
\newblock {\em JETP Letters}, 102(10):650--654.

\bibitem[Ruban, 2015b]{ruban2015b}
Ruban, V.~P. (2015b).
\newblock Gaussian variational ansatz in the problem of anomalous sea waves:
  {C}omparison with direct numerical simulation.
\newblock {\em Journal of Experimental and Theoretical Physics},
  120(5):925--932.

\bibitem[Rudin, 1964]{rudin64}
Rudin, W. (1964).
\newblock {\em Principles of mathematical analysis}, volume~3.
\newblock McGraw-Hill New York.

\bibitem[Shemer et~al., 2010]{shemer2010}
Shemer, L., Sergeeva, A., and Liberzon, D. (2010).
\newblock Effect of the initial spectrum on the spatial evolution of statistics
  of unidirectional nonlinear random waves.
\newblock {\em Journal of Geophysical Research: Oceans}, 115(C12).

\bibitem[Story et~al., 2011]{story11}
Story, W.~R., Fu, T.~C., and Hackett, E.~E. (2011).
\newblock Radar measurement of ocean waves.
\newblock In {\em ASME 2011 30th International Conference on Ocean, Offshore
  and Arctic Engineering}, pages 707--717.

\bibitem[Tayfun, 1980]{tayfun1980}
Tayfun, M.~A. (1980).
\newblock Narrow-band nonlinear sea waves.
\newblock {\em J. Geophys. Res.}, 85(C3):1548--1552.

\bibitem[Trillo et~al., 2016]{Trillo16}
Trillo, S., Deng, G., Biondini, G., Klein, M., Clauss, G.~F., Chabchoub, A.,
  and Onorato, M. (2016).
\newblock Experimental observation and theoretical description of multisoliton
  fission in shallow water.
\newblock {\em Phys. Rev. Lett.}, 117:144102.

\bibitem[Trulsen and Dysthe, 1996]{trulsen96}
Trulsen, K. and Dysthe, K.~B. (1996).
\newblock A modified nonlinear {S}chr{\"o}dinger equation for broader bandwidth
  gravity waves on deep water.
\newblock {\em Wave motion}, 24(3):281--289.

\bibitem[Trulsen et~al., 2000]{trulsen2000}
Trulsen, K., Kliakhandler, I., Dysthe, K.~B., and Velarde, M.~G. (2000).
\newblock On weakly nonlinear modulation of waves on deep water.
\newblock {\em Phys. Fluids}, 12(10):2432--2437.

\bibitem[Trulsen and Stansberg, 2001]{trulsen01}
Trulsen, K. and Stansberg, C.~T. (2001).
\newblock Spatial evolution of water surface waves: {N}umerical simulation and
  experiment of bichromatic waves.
\newblock In {\em The Eleventh International Offshore and Polar Engineering
  Conference}.

\bibitem[Tulin and Waseda, 1999]{tulin1999}
Tulin, M.~P. and Waseda, T. (1999).
\newblock Laboratory observations of wave group evolution, including breaking
  effects.
\newblock {\em J. Fluid Mech.}, 378:197--232.

\bibitem[Xiao, 2013]{xiaoThesis}
Xiao, W. (2013).
\newblock {\em Study of directional ocean wavefield evolution and rogue wave
  occurrence using large-scale phase-resolved nonlinear simulations}.
\newblock PhD thesis, Massachusetts Institute of Technology.

\bibitem[Xiao et~al., 2013]{xiao2013}
Xiao, W., Liu, Y., Wu, G., and Yue, D. K.~P. (2013).
\newblock Rogue wave occurrence and dynamics by direct simulations of nonlinear
  wave-field evolution.
\newblock {\em J. Fluid Mech.}, 720:357--392.

\bibitem[Zakharov, 1968]{zakharov68}
Zakharov, V.~E. (1968).
\newblock Stability of periodic waves of finite amplitude on the surface of a
  deep fluid.
\newblock {\em Journal of Applied Mechanics and Technical Physics},
  9(2):190--194.
\end{thebibliography}

\end{document}